\documentclass[]{aastex631}

\usepackage{amsmath}
\usepackage[ruled,vlined]{algorithm2e}
\usepackage{rotating}
\usepackage{hyperref}
\usepackage{placeins}
\submitjournal{AJ}

\shorttitle{Mass ratio of single-line visual binaries}
\shortauthors{Anguita-Aguero et al.}
\graphicspath{{./}{figures/}}

\begin{document}

\title{Mass ratio of single-line spectroscopic binaries with visual
  orbits using Bayesian inference and suitable priors\footnote{Based
    in part on observations obtained at the Southern Astrophysical
    Research (SOAR) telescope, which is a joint project of the
    Minist\'{e}rio da Ci\^{e}ncia, Tecnologia, e Inova\c{c}\~{a}oes
    (MCTI) da Rep\'{u}blica Federativa do Brasil, the U.S. National
    Optical Astronomy Observatory (NOAO), the University of North
    Carolina at Chapel Hill (UNC), and Michigan State University
    (MSU).}}

\correspondingauthor{Rene A. Mendez}
\email{rmendez@uchile.cl}

\author{Jennifer Anguita-Aguero}
\affiliation{Astronomy Department \\ Universidad de Chile \\ Casilla 36-D, Santiago, Chile}

\author[0000-0003-1454-0596]{Rene A. Mendez}
\affiliation{Astronomy Department \\ Universidad de Chile \\ Casilla 36-D, Santiago, Chile\\
and \\
European Southern Observatory \\
Alonso de Córdova 3107, Vitacura\\
Santiago, Chile}

\author[0000-0002-7140-0437]{Miguel Videla}
\affiliation{Department of Electrical Engineering\\
Information and Decision Systems Group (IDS) \\
Facultad de Ciencias Físicas y Matemáticas, Universidad de Chile \\
Beauchef 850, Santiago, Chile}
  
\author{Edgardo Costa}
\affiliation{Astronomy Department \\ Universidad de Chile \\ Casilla 36-D, Santiago, Chile}

\author{Leonardo Vanzi}
\affiliation{Department of Electrical Engineering and Centre of Astro Engineering \\ Pontificia Universidad Católica de Chile \\ Av. Vicuña Mackenna 4860, Santiago}
 
\author{Nicolas Castro-Morales}
\affiliation{Instituto de Astrofísica \\ Pontificia Universidad Católica de Chile \\ Instituto Milenio de Astrofísica (MAS) \\ Av. Vicuña Mackenna 4860, Santiago}

\author{Camila Caballero-Valdes}
\affiliation{Astronomy Department \\ Universidad de Chile \\ Casilla 36-D, Santiago, Chile}


\begin{abstract}
We present orbital elements for twenty-two single-line binaries, nine
of them studied for the first time, determined from a joint
spectroscopic and astrometric solution. The astrometry is based on
interferometric measurements obtained with the HRCam Speckle camera on
the SOAR 4.1m telescope at Cerro Pachon, Chile, supplemented with
historical data. The spectroscopic observations were secured using
Echelle spectrographs (FEROS, FIDEOS and HARPS) at La Silla, Chile. A
comparison of our orbital elements and systemic velocities with
previous studies, including Gaia radial velocities, show the
robustness of our estimations. By adopting suitable priors of the
trigonometric parallax and spectral type of the primary component, and
using a Bayesian inference methodology developed by our group, we were
able to estimate mass ratios for these binaries. Combining the present
results with a previous study of other single-line from our team we
present a pseudo mass-to-luminosity relationship based on twenty three
systems (45 stars) in the mass range $0.6 \le M_\odot \le 2.5$. We
find a reasonable correspondence with a fiducial mass-to-luminosity
relationship. We conclude that our methodology does allow to derive
tentative mass ratios for this type of binaries.
\end{abstract}




\section{Introduction} \label{sec:intro}
Spectroscopic binaries are powerful astrophysical laboratories. Combining precise astrometric and radial velocity (RV) measurements for these systems it is possible to obtain a fairly complete characterization of their orbital and basic astrophysical parameters. Among them, two groups are distinguished: double-line spectroscopic binaries ($SB2$), in whose spectra the spectral lines of both components are distinguished, and single-line spectroscopic binaries ($SB1$), for which only the lines of the primary component are easily recognized.

$SB2$ are certainly the most interesting systems because in their case a joint treatment of the atrometric and RV data allows to determine directly the individual component masses (\citet{2022AJ....163..118A}), as well as a parallax -free distance- allowing an independent assessment of Gaia’s trigonometric parallaxes (\citet{Pour2000}, \citet{Mas2015}).

Unfortunately, for $SB1s$ -which are the majority (67\%) of the systems included in the 9th Catalog of Spectroscopic Binary Orbits\footnote{Updated regularly, and available at \url{https://sb9.astro.ulb.ac.be/}.}, \citet{Pouret2004}, SB9, only the mass function can be obtained directly\footnote{The mass function is defined by $f(M)=\frac{\left( m_{\mbox{s}} \sin i \right) ^3}{\left( m_{\mbox{p}} + m_{\mbox{s}} \right) ^2}$, where $m_{\mbox{p}}$ and $m_{\mbox{s}}$ are the mass of the primary and secondary stars respectively} (\citet{Struve1958}). For this reason, in the past this latter group has not been fully exploited.

Now, thanks to a newly developed Bayesian methodology based on the MCMC algorithm, No-U-Turn sampler \citep{ViMe22} to address the orbital parameters inference problem in $SB1$ systems, including an estimation of the individual component masses, this situation is rapidly changing. This scheme also provides a precise characterization of the uncertainty of the estimates of the orbital parameters, in the form of joint posterior probability distribution function.

In this approach, the lack of an RV curve for the secondary star is managed incorporating suitable prior distributions for two critical parameters of the system; namely its trigonometric parallax (from an external source), and the mass of the primary component, estimated from its Spectral Type (hereafter  SpTy). This methodology has been thoroughly tested on several benchmark $SB2$ systems by \citet{ViMe22}, who provide an exhaustive analysis of the results obtained by comparing the PDFs from different observational sources and priors. In that paper we were able to show that this Bayesian approach allows a much richer and complete understanding of the associated uncertainties in the study of binary systems in general.

Here we apply this methodology to twenty two $SB1$ systems, nine of which (HIP \# 29860, 36497, 38414, 40167, 54061, 76031, 93017, 96302, and 116259)
up to now did not have a published self-consistent spectroscopic/astrometric joint orbital solution.

The paper is organized as follows: in Section~\ref{sec:sample} we present our list of $SB1s$, together with basic information relevant for our study; in Section~\ref{sec:oel} we present the results of our orbital calculations and mass ratios, and in Section~\ref{sec:objnotes} we present a detailed discussion of each of our objects. Finally, in Section~\ref{sec:concl} we present the main conclusions of our study.



\section{The \textit{SB1}  sample}
\label{sec:sample}

To select the sample for the present work, we started by doing a cross-match between the Sixth Catalog of Orbits of Visual Binary Stars maintained by the US Naval Observatory (hereafter Orb6\footnote{Available at \url{https://www.usno.navy.mil/USNO/astrometry/optical-IR-prod/wds/orb6}}) and SB9. 

Orb6 is the most comprehensive catalog of binary systems with published orbital elements, while SB9 contains RV amplitudes for all binary systems for which it has been possible to fit a RV curve. Having identified those systems confirmed as $SB1$ in SB9, we pinpointed the binaries for which a combined astrometric/RV study of the orbit was not available in the literature by means of the notes and comments given in Orb6 and SB9, or which merit further study given new available data. This lead to an initial working list of thirty six binary systems.

For the systems selected as indicated above, we retrieved their RV data from SB9, or from references provided therein; and the astrometric data from the US Naval Observatory Fourth Catalog of Interferometric Measurements of Binary Stars\footnote{The latest version, called int4, is available at \url{https://www.usno.navy.mil/USNO/astrometry/optical-IR-prod/wds/int4/fourth-catalog-of-interferometric-measurements-of-binary-stars}} and from historical astrometry included in the Washington Double Star Catalog effort (\citet{WDSCat2001}, hereafter WDS, kindly provided to us by Dr.Brian Mason from the US Naval Observatory).

Finally, we included recent results obtained with the HRCam Speckle camera on the SOAR 4.1m telescope at Cerro Pachón\footnote{For up-to-date details of the instrument see \url{http://www.ctio.noao.edu/~atokovin/speckle/}},
as part of our monitoring of Southern binaries described in \citet{Mendezetal2017}.
We note that some of these measurements have been secured after the publication of their last orbit, which allows for an improvement of the orbital solutions. We have also supplemented the published RV with our own recent observations secured with the FEROS Echelle \citep{1999Msngr..95....8K}  high-resolution spectrograph on the 2.2m MPG telescope\footnote{See \url{https://www.eso.org/sci/facilities/lasilla/instruments/feros.html}}, and the FIDEOS Echelle \citep{2018MNRAS.477.5041V} on the 1m telescope\footnote{See \url{https://www.eso.org/public/teles-instr/lasilla/1metre/fideos/}}, both operating at the ESO/La Silla Observatory, Chile. FEROS and FIDEOS spectra were reduced using the CERES pipeline \citep{2017PASP..129c4002B}. In a couple of cases we also found high-precision RV archival data for our binaries obtained during the planet-monitoring program being carried out with HARPS \citep{2003Msngr.114...20M} on the 3.6m telescope at ESO/La Silla\footnote{See \url{https://www.eso.org/sci/facilities/lasilla/instruments/harps.html}}. This added valuable and highly precise points to the RV curve.

Examination of the information collected showed that only thirty four of the systems in our starting list had sufficient data to warrant further analysis. Out of this final working list, twelve $SB1$ systems were presented and studied in \citet{ViMe22}, while the remaining twenty-two are included in the present paper. We must emphasize that, as a result of our selection process, our final sample is very heterogeneous and it should not be considered complete or representative of $SB1$ systems in any astrophysical sense. From this point of view, the main contribution of this paper is the addition of new orbits and mass ratios for this type of binaries, nine of which do not have a published joint estimation of their orbital parameters (to the best of our knowledge, ours is therefore the first combined orbit).


Table~\ref{tab:SB1/SB1_plx_m1} presents basic properties available in the literature for the sample studied in this work. The first two columns present the Hipparcos number and the discovery designation code assigned in the WDS Catalog. The following columns present the trigonometric parallax adopted as prior, the RUWE (Reduced Unit Weight Error - an indication of the reliability of the parallax) parameter as given in the Gaia catalogue, the SpTy adopted for the primary component (from SIMBAD, \citet{SIMBAD}, WDS, or our own estimate - as explained below), and the mass of the primary component implied by the SpTy. 

The masses have been derived from the mass versus SpTy and mass versus luminosity class calibrations, provided by \citet{Abuset2020} or, if not available there, from \citet{1981Ap&SS..80..353S}. The dispersion in mass comes from assuming a SpTy uncertainty of $\pm$ one sub-type, which is customary in spectral classification. As it can be seen from this table, and mentioned above, our $SB1s$ represent an heterogeneous group of binaries, with masses in the range between 0.4$M_\odot$ to slightly above 6$M_\odot$, located at distances between 7 to 263~pc. Also, as we shall see in the following Sections, the data quality and orbital phase coverage available for sample is quite varied.


\small
\begin{table}[h]
\caption{Trigonometric parallax, SpTy (primary component) and mass (primary component) of the $SB1$ stellar systems presented in this paper. See text for details. \label{tab:SB1/SB1_plx_m1}}
\begin{tabular}{cccccc}
\hline\hline
HIP \# & Discovery   & $\varpi$\tablenotemark{a}   &
RUWE\tablenotemark{a} & SpTy\tablenotemark{b} & $m_1$\tablenotemark{c}  \\
       & Designation & [mas] &      &  & [M$_\odot]$ \\
\hline
3850 & PES1 &$53.053\pm 0.028$  & 0.9325 &  G9V\tablenotemark{f} &   $0.93\pm0.04 $ \\
5336 & WCK1Aa,Ab & $130.29\pm0.44$ & 6.9658 &  G5V &   $1.05\pm0.04$ \\
17491 & BAG8AB  &$40.33\pm 0.25$ & 11.5927 & K0V &   $0.89\pm0.04$ \\
28691 & MCA24   &$3.82\pm 0.25$ & 2.5179 & B8III &   $4.26\pm1.15$ \\
29860 & CAT1Aa,Ab &$51.62\pm 0.12$ & 1.9793 & F9V/G0V\tablenotemark{f} &   $1.175\pm0.025$ \\
36497 & TOK392Da,Db &$21.19\pm 0.18$ & 5.8032 & F8V &   $1.23\pm0.05$ \\
38414 & TOK195 &$8.98\pm 0.23$ & 6.0363 & K1/2II\tablenotemark{f}   &   $6.09\pm0.07$\tablenotemark{g} \\
39261 & MCA33 & $10.24\pm 0.22$ & 5.8681 & A2V/A3V\tablenotemark{f}  &   $2.32\pm0.08$ \\
40167 & HUT1Ca,Cb & $40.89\pm 0.15$ & 1.3840 & F8V  &   $1.23\pm0.05$ \\
43109 & SP1AB & $26.437\pm 0.098$ & 1.4190 & G1III  &   $1.02\pm0.20$ \\
54061 & BU1077AB & $26.54\pm 0.48$\tablenotemark{d} & -- & G9III  &   $1.93\pm0.26$ \\
55642 & STF1536AB & $41.93\pm 0.43$\tablenotemark{e,h} & -- & F4IV  &   $1.380\pm0.035$ \\
67620 & WSI77 & $53.88\pm 0.34$\tablenotemark{e} & -- & G5V  &   $1.05\pm0.04$ \\
75695 & JEF1  & $29.17\pm 0.76$\tablenotemark{d} & 7.2964 & A5V  &   $2.00\pm0.06$ \\
76031 & TOK48  & $19.67\pm 0.89$\tablenotemark{d} & -- & G0V  &   $1.15\pm0.04$ \\
78727 & STF1998AB & $35.89 \pm 0.23$\tablenotemark{i} & 1.2934 & F5IV  &   $1.350\pm0.025$ \\
93017 & BU648AB & $67.14\pm 0.12$ & 2.7280 & F9V/G0V\tablenotemark{f}  &   $1.175\pm0.025$ \\
96302 & WRH32 & $5.37\pm 0.10$ & 1.5265 & G8III  &   $1.87\pm0.26$ \\
103655 & KUI103 & $66.554\pm 0.072$ & 5.2017 & M2V  &   $0.43\pm0.02$ \\
111685 & HDS3211AB  & $51.2\pm 1.6$\tablenotemark{d} & 31.988 & M0V &   $0.54\pm0.02$ \\
111974 & HO296AB  & $29.59\pm 0.68$\tablenotemark{d} & -- & G4V  &   $1.06\pm0.04$ \\
116259 & HDS3356 & $29.23\pm 0.15$ & 8.0906 & G0V  &   $1.15\pm0.04$ \\
\hline\hline
\end{tabular}
\tablenotetext{a}{From GAIA DR3 except when noted}
\tablenotetext{b}{From SIMBAD, WDS or from our own estimate, see text for details}
\tablenotetext{c}{From \citet{Abuset2020} except when noted}
\tablenotetext{d}{From HIPPARCOS}
\tablenotetext{e}{From GAIA DR2}
\tablenotetext{f}{Adopted the mean spectral type}
\tablenotetext{g}{From \citet{1981Ap&SS..80..353S}}
\tablenotetext{h}{Average of Gaia DR2 parallaxes for AB and C components}
\tablenotetext{i}{Average of Gaia DR3 parallaxes for AB and C components}
\end{table}
\normalsize

Regarding the trigonometric parallaxes used as priors, and indicated in the third column of Table~\ref{tab:SB1/SB1_plx_m1}, we note that for unresolved binary systems (separations smaller than about 0.7~arcsec) and multiple systems, the Gaia solution can be compromised by acceleration and/or unresolved companions because the current astrometric reductions assume single stars. The RUWE parameter given in the fourth column of this table highlights excessive astrometric noise, helping to identify suspicious astrometry (mainly those with RUWE $> 2.0$; see, e.g., \citet{2022AJ....163..161T}). Accordingly, as can be seen in this table, most (but not all) of the systems studied here indeed have a high RUWE. This is an important issue that must be considered when analyzing the results for individual objects, and the consistency between the different solutions.

To assign the SpTy to the primary components, listed in column five of Table~\ref{tab:SB1/SB1_plx_m1}, we consulted SIMBAD, the WDS catalog itself, and the Catalogue of Stellar Spectral Classifications by \citet{2014yCat....102023S}, which provides a compilation  of spectral classifications determined from spectral data alone (i.e., no narrow-band photometry), and which is updated regularly in VizieR (catalog B/mk/mktypes, currently containing more than 90.000 stars).

A comparison of the data from these three sources revealed that some objects in our sample have somewhat ambiguous SpTy. In order to resolve these ambiguities, we computed the absolute magnitude of the primary component using the trigonometric parallax of the system listed in Table~\ref{tab:SB1/SB1_plx_m1} and their apparent magnitudes given in the WDS catalogue. This absolute magnitude was then compared with the absolute magnitude expected from the listed SpTy, using the calibrations provided by \citet{Abuset2020}). The SpTy closest to the computed absolute magnitude was finally adopted. We note that some ambiguities in the SpTy persisted after these calculations (see Section~\ref{sec:objnotes}).

\newpage
\section{Orbital elements and mass ratios} \label{sec:oel}

To determine the orbital parameters, we followed the scheme presented in detail in \citet{ViMe22}. In summary, for each object, we are able to compute four orbital solutions, namely: one using no priors as in classical works, denoted as $SB1$ solution; one using the trigonometric parallax as a prior, denoted $SB1+p(\varpi)$ solution; one using the mass of the primary as a prior, denoted $SB1+p(m_1|\theta)$ solution; and finally, a combined solution using parallax and mass as priors simultaneously, denoted $SB1+p(\varpi)+p(m_1|\theta)$. These solutions are presented in Table~\ref{tab:SB1/SB1}. For reference, in the first two lines of each entry we also present the orbital parameters given in Orb6 and SB9.

For each of our solutions, we provide the orbital elements obtained from the Maximum A-Posteriori estimation (MAP), which gives the most probable sample of the PDF, as well as the upper and lower values that encompass the $95\%$ credible interval around the MAP solution (denoted HDPI, for High Posterior Density Interval and which encompasses the mode).

\begin{figure}[h]
\includegraphics[width=\textwidth]{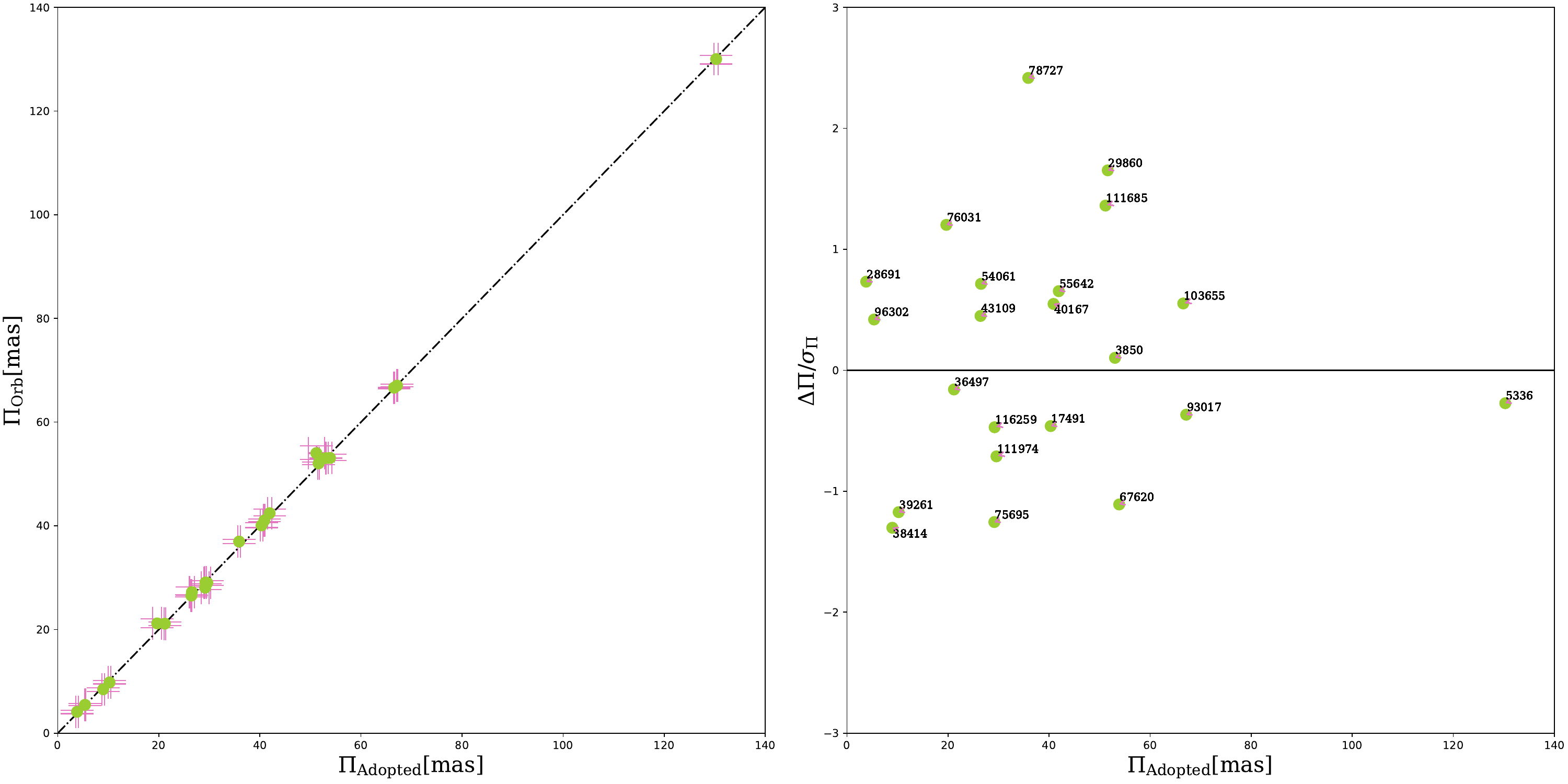}
\caption{(pseudo) Orbital parallaxes from this work, versus the adopted prior parallax (left panel). In the right panel we show the residuals in the sense $\Pi_{\mbox{Orb}} - \Pi_{\mbox{Adopted}}$ normalized by the parallax uncertainty of each target (see text for details). The labels indicate the HIP number.
\label{fig:plx_sb1}}
\end{figure}

Of course, as explained in detail in \citet{ViMe22}, only solutions that include a prior can lead to an estimation of the orbital parallax and the mass ratio, which are presented in the last columns of Table~\ref{tab:SB1/SB1}. We must note that the inferred parallaxes reported in this table cannot be properly called orbital parallaxes in the classical sense (as in the case of $SB2$s systems), because, while they have been derived self-consistently from the model and data, they are only resolvable by the incorporation of the priors. Nevertheless, throughout this paper we will still refer to these as  orbital parallaxes, to differentiate them from the trigonometric parallax adopted as prior. A plot of our MAP (pseudo) orbital parallaxes (from Table~\ref{tab:SB1/SB1}) versus the adopted prior parallax from Table~\ref{tab:SB1/SB1_plx_m1} is shown in Figure~\ref{fig:plx_sb1}, which exhibits a general good agreement between them, with a global rms of 0.97~mas over our 22 objects. In the right panel on Figure~\ref{fig:plx_sb1} the residuals have been normalized by the overall parallax uncertainty, which includes the uncertainty in the adopted parallax added in quadrature to the uncertainty of our estimated parallax.

\FloatBarrier
\FloatBarrier

A look at the results on Table~\ref{tab:SB1/SB1} shows that our values for the orbital elements in general coincide  quite well with those from previous studies. In particular, it is well known that periapsis ($\omega$) can only be well-determined by RV measurements as long as the distinction between primary and secondary is unambiguous (which is difficult, e.g., for equal-mass binaries); and the table shows our values are indeed quite close to those from SB9, but with smaller uncertainties in our case\footnote{Two notable cases where we maintain the $\omega$ value from Orb6, at odds with that reported from SB9 have indeed a $q\sim1$, namely HIP~78727 and HIP~111974.}. 

On the other hand, the longitude of the ascending node ($\Omega$) can be well determined from astrometric observations alone, but it is ambiguous in the case of equal-brightness binaries; and, additionally, it is subject to an ambiguity of $\pm$180$^{\circ}$, in which case the $\omega$ value is also affected by the same ambiguity (see Equation (28) in Appendix~B.2 of \citet{Mendezetal2017}).

This is clearly seen on HIP~28691: subtracting 180$^{\circ}$ from $\Omega$ and $\omega$ on Orb6 gives SB9's and our values and on HIP~43109 and HIP~54061 where, now, adding 180$^{\circ}$ from $\Omega$ and $\omega$ on Orb6 gives SB9's and our values. Apart from these two cases, from the table we see that there is good correspondence between  our values for $\Omega$ and those from Orb6 (but again with smaller formal uncertainties in our case, with a few exceptions). In terms of the other orbital elements, despite the fact that the sample is very heterogeneous, e.g., with periods ranging from 1.7~yr (HIP~76031) to 189~yrs (HIP~55642) and with semi-major axis from 25~mas (HIP~96302) to 1.9~arcsec (HIP~55642) our solutions are, again, similar to those of previous studies, with the notable exception of HIP~29860, where a large difference is seen between our solution and previous studies. This case is further described below, and in Figures~\ref{fig:29860_3} and \ref{fig:29860_1}. More specific notes and comments on individual objects are given below, in Section~\ref{sec:objnotes}.

\FloatBarrier
\small
\begin{sidewaystable}[ht]
\centering
\caption{Table 2. MAP estimates and 95\% HDPIs derived from the marginal PDFs of the orbital parameters for the four orbital solutions, based on our astrometric data, discussed in the first paragraph of Section 3. $m_1$ and $\varpi$ are the priors used for the mass of the primary component and the trigonometric parallax, respectively. In the first two lines (of six entries) for each $SB1$ system, we report the values provided by Orb6 and SB9, preserving the significant figures included in those catalogues as an indication of their precision (continued on next page). \label{tab:SB1/SB1}}
\resizebox{\textwidth}{!}{\begin{tabular}{cccccccccccccc}
\hline\hline
    HIP~\# &                     System &                 $P$ &                       $T$ &                     $e$ &              $a$ &                   $\omega$  &                $\Omega$ &                     $i$  &                $V_0$&              $\varpi$ &                $f/\varpi$\tablenotemark{a} &         $m_1$  &                     $q$ \\
 & & [yr] & [yr] &  & [arcsec] &   [\textdegree] &  [\textdegree] & [\textdegree] & [km/s] & [mas] & [pc] &  [M$_\odot]$ & \\
\hline
3850 & Orb6 & $33.08 \pm 0.70$ & $1997.066 \pm 0.0143$ & $0.732 \pm 0.0023$ & $0.5199 \pm 0.0512$ & $266.90 \pm 0.47$ & $89.9 \pm 1.4$ & $46.3 \pm 1.1$ & --- & --- & --- & --- & --- \\
 & SB9 & $31.74 \pm 3.0626$ & $2028.7857 \pm 3.068$ & $0.723 \pm 0.013$ & --- & $-94.2 \pm 1.6$ & --- & --- & $9.93 \pm Fixed$ & --- & --- & --- & --- \\
         &                      $SB1$ & $41.235_{36.471}^{48.786}$ & $2032.630_{2030.426}^{2035.724}$ & $0.456_{0.363}^{0.524}$ & $0.426_{0.410}^{0.451}$ & $331.331_{318.163}^{344.740}$ & $17.371_{2.360}^{32.777}$ &    $4.529_{0.380}^{15.426}$ & $10.325_{10.162}^{10.574}$ &                         --- &       $24.804_{3.693}^{173.460}$ &                         --- &                       --- \\
          &               $SB1+p(\varpi)$ & $31.754_{29.581}^{35.091}$ & $2029.636_{2027.474}^{2032.741}$ & $0.758_{0.737}^{0.774}$ & $0.543_{0.483}^{0.624}$ & $263.688_{262.780}^{267.484}$ & $94.955_{84.513}^{103.084}$ &    $50.298_{43.456}^{58.591}$ & $9.918_{9.880}^{9.962}$ & $53.057_{52.999}^{53.106}$ &       $1.155_{1.001}^{1.327}$ &   $1.001_{0.777}^{1.251}$ & $0.065_{0.056}^{0.076}$ \\
          &        $SB1+p(m_1|\theta)$ & $39.291_{36.457}^{49.103}$ & $2031.769_{2030.320}^{2035.702}$ & $0.480_{0.349}^{0.525}$ & $0.421_{0.410}^{0.456}$ & $326.008_{318.748}^{344.392}$ & $22.870_{0.017}^{29.713}$ &    $7.705_{4.913}^{19.077}$ & $10.262_{10.171}^{10.600}$ & $31.556_{27.870}^{34.916}$ &       $12.609_{6.263}^{17.658}$ &   $0.926_{0.848}^{1.003}$ & $0.661_{0.296}^{1.000}$ \\
          & $SB1+p(\varpi)+p(m_1|\theta)$ & $31.305_{29.895}^{32.926}$ & $2029.210_{2028.028}^{2030.483}$ & $0.753_{0.742}^{0.762}$ & $0.528_{0.503}^{0.555}$ & $263.981_{263.004}^{266.640}$ & $93.501_{86.936}^{96.745}$ &    $48.460_{45.311}^{51.757}$ & $9.926_{9.908}^{9.951}$ & $53.059_{52.996}^{53.107}$ &       $1.189_{1.115}^{1.297}$ &   $0.941_{0.864}^{1.012}$ & $0.067_{0.063}^{0.074}$ \\
\hline
5336 & Orb6 & $21.568 \pm 0.015$ & $1997.2235 \pm 0.0067$ & $0.5885 \pm 0.0011$ & $0.9985 \pm 0.0013$ & $330.37 \pm 0.18$ & $223.868 \pm 0.064$ & $110.671 \pm 0.064$ & --- & --- & --- & --- & --- \\
 & SB9 & $21.4 \pm 0.11$ & $1975.901 \pm 0.1$ & $0.53 \pm 0.02$ & --- & $150.0 \pm 3.4$ & --- & --- & $-97.35 \pm 0.04$ & --- & --- & --- & --- \\
         &                      $SB1$ & $21.289_{21.250}^{21.317}$ & $1997.358_{1997.279}^{1997.434}$ & $0.558_{0.551}^{0.565}$ & $1.029_{1.023}^{1.035}$ & $151.763_{150.371}^{153.274}$ & $45.581_{45.202}^{46.037}$ &    $107.723_{107.483}^{107.953}$ & $-97.413_{-97.526}^{-97.287}$ &                         --- &       $1.322_{1.233}^{1.436}$ &                         --- &                       --- \\
          &               $SB1+p(\varpi)$ & $21.290_{21.251}^{21.319}$ & $1997.360_{1997.278}^{1997.433}$ & $0.558_{0.552}^{0.565}$ & $1.028_{1.023}^{1.035}$ & $151.854_{150.354}^{153.306}$ & $45.604_{45.202}^{46.053}$ &    $107.723_{107.472}^{107.941}$ & $-97.371_{-97.519}^{-97.281}$ & $130.343_{129.468}^{131.162}$ &       $1.371_{1.246}^{1.445}$ &   $0.889_{0.868}^{0.927}$ & $0.218_{0.193}^{0.232}$ \\
          &        $SB1+p(m_1|\theta)$ & $21.280_{21.251}^{21.316}$ & $1997.347_{1997.276}^{1997.429}$ & $0.558_{0.552}^{0.565}$ & $1.029_{1.023}^{1.035}$ & $151.683_{150.304}^{153.196}$ & $45.599_{45.201}^{46.031}$ &    $107.745_{107.474}^{107.948}$ & $-97.397_{-97.520}^{-97.286}$ & $124.211_{121.247}^{127.363}$ &       $1.360_{1.239}^{1.444}$ &   $1.043_{0.966}^{1.123}$ & $0.203_{0.182}^{0.218}$ \\
          & $SB1+p(\varpi)+p(m_1|\theta)$ & $21.282_{21.246}^{21.312}$ & $1997.335_{1997.261}^{1997.417}$ & $0.558_{0.551}^{0.565}$ & $1.032_{1.025}^{1.037}$ & $151.388_{150.040}^{152.982}$ & $45.502_{45.125}^{45.973}$ &    $107.667_{107.411}^{107.889}$ & $-97.454_{-97.542}^{-97.301}$ & $130.034_{129.081}^{130.732}$ &       $1.271_{1.210}^{1.412}$ &   $0.921_{0.889}^{0.946}$ & $0.198_{0.186}^{0.224}$ \\
\hline
17491 & Orb6 & $13.877 \pm 0.016$ & $1992.092 \pm 0.016$ & $0.2764 \pm 0.003$ & $0.2574 \pm 0.0046$ & $74.85 \pm 0.25$ & $281.1 \pm 1.0$ & $101.37 \pm 0.12$ & --- & --- & --- & --- & --- \\
 & SB9 & $14.021 \pm 0.123$ & $1992.039 \pm 0.070$ & $0.275 \pm 0.010$ & --- & $72.30 \pm 2.16$ & --- & --- & $22.238 \pm 0.044$ & --- & --- & --- & --- \\
         &                      $SB1$ & $13.762_{13.658}^{13.897}$ & $1992.017_{1991.916}^{1992.111}$ & $0.263_{0.247}^{0.280}$ & $0.254_{0.249}^{0.258}$ & $72.003_{69.482}^{74.567}$ & $281.083_{279.943}^{281.991}$ &    $101.987_{101.148}^{102.718}$ & $22.312_{22.183}^{22.390}$ &                         --- &       $10.330_{9.979}^{10.708}$ &                         --- &                       --- \\
          &               $SB1+p(\varpi)$ & $13.767_{13.656}^{13.897}$ & $1992.012_{1991.910}^{1992.108}$ & $0.265_{0.248}^{0.281}$ & $0.253_{0.249}^{0.258}$ & $71.985_{69.425}^{74.497}$ & $280.834_{279.916}^{281.952}$ &    $102.018_{101.168}^{102.723}$ & $22.288_{22.175}^{22.384}$ & $40.215_{39.819}^{40.800}$ &       $10.296_{10.019}^{10.750}$ &   $0.773_{0.698}^{0.827}$ & $0.707_{0.672}^{0.764}$ \\
          &        $SB1+p(m_1|\theta)$ & $13.801_{13.653}^{13.897}$ & $1992.020_{1991.918}^{1992.113}$ & $0.264_{0.247}^{0.280}$ & $0.253_{0.249}^{0.258}$ & $72.409_{69.625}^{74.630}$ & $281.113_{279.850}^{281.932}$ &    $101.778_{101.141}^{102.680}$ & $22.306_{22.179}^{22.387}$ & $38.620_{37.553}^{39.907}$ &       $10.395_{10.001}^{10.728}$ &   $0.884_{0.809}^{0.965}$ & $0.671_{0.633}^{0.707}$ \\
          & $SB1+p(\varpi)+p(m_1|\theta)$ & $13.721_{13.638}^{13.879}$ & $1992.035_{1991.933}^{1992.126}$ & $0.274_{0.254}^{0.286}$ & $0.257_{0.252}^{0.260}$ & $72.234_{69.683}^{74.703}$ & $280.936_{279.929}^{281.963}$ &    $101.624_{100.973}^{102.474}$ & $22.267_{22.173}^{22.384}$ & $40.081_{39.651}^{40.591}$ &       $10.117_{9.833}^{10.461}$ &   $0.828_{0.766}^{0.869}$ & $0.682_{0.652}^{0.724}$ \\
\hline
28691 & Orb6 & $13.061 \pm 0.03$ & $2006.52 \pm 0.16$ & $0.808 \pm 0.045$ & $0.0555 \pm 0.0022$ & $291.5 \pm 9.2$ & $224.4 \pm 3.3$ & $111.8 \pm 7.8$ & --- & --- & --- & --- & --- \\
 & SB9 & $12.98 \pm 0.014$ & $1980.807 \pm 0.015$ & $0.7407 \pm 0.0042$ & --- & $125.30 \pm 0.71$ & --- & --- & $12.607 \pm 0.052$ & --- & --- & --- & --- \\
         &                      $SB1$ & $13.012_{12.975}^{13.040}$ & $1928.728_{1928.590}^{1928.917}$ & $0.739_{0.730}^{0.747}$ & $0.046_{0.044}^{0.048}$ & $125.335_{124.013}^{126.837}$ & $51.094_{47.951}^{55.451}$ &    $116.340_{110.006}^{122.285}$ & $12.607_{12.491}^{12.710}$ &                         --- &       $86.624_{79.458}^{94.033}$ &                         --- &                       --- \\
          &               $SB1+p(\varpi)$ & $13.006_{12.977}^{13.040}$ & $1928.760_{1928.588}^{1928.906}$ & $0.737_{0.730}^{0.747}$ & $0.046_{0.044}^{0.048}$ & $125.758_{123.964}^{126.881}$ & $52.315_{48.112}^{55.542}$ &    $116.431_{110.387}^{122.709}$ & $12.580_{12.489}^{12.708}$ & $3.867_{3.333}^{4.316}$ &       $86.960_{79.925}^{94.938}$ &   $6.543_{3.917}^{10.555}$ & $0.507_{0.391}^{0.623}$ \\
          &        $SB1+p(m_1|\theta)$ & $13.009_{12.976}^{13.041}$ & $1928.749_{1928.593}^{1928.916}$ & $0.737_{0.730}^{0.747}$ & $0.046_{0.044}^{0.048}$ & $125.603_{124.043}^{126.919}$ & $52.137_{48.022}^{55.530}$ &    $116.018_{110.283}^{122.700}$ & $12.632_{12.491}^{12.708}$ & $4.246_{3.853}^{5.528}$ &       $86.694_{79.729}^{94.788}$ &   $4.727_{1.497}^{5.790}$ & $0.582_{0.493}^{0.901}$ \\
          & $SB1+p(\varpi)+p(m_1|\theta)$ & $13.013_{12.979}^{13.043}$ & $1928.727_{1928.584}^{1928.905}$ & $0.738_{0.730}^{0.748}$ & $0.046_{0.044}^{0.047}$ & $125.530_{124.055}^{126.928}$ & $51.655_{47.300}^{54.833}$ &    $116.000_{111.186}^{123.755}$ & $12.606_{12.487}^{12.706}$ & $4.122_{3.756}^{4.433}$ &       $85.971_{81.065}^{96.688}$ &   $5.334_{3.785}^{6.887}$ & $0.549_{0.477}^{0.665}$ \\          
\hline
29860 & Orb6 & 35.52 & 1998.057 & 0.809 & 0.648 & 76.0 & 166.8 & 41.3 & --- & --- & --- & --- & --- \\
 & SB9 & $34.20 \pm 0.234$ & $1998.050 \pm 0.0013$ & $0.8045 \pm 0.0009$ & --- & $75.21 \pm 0.15$ & --- & --- & $-10.035 \pm Fixed$ & --- & --- & --- & --- \\
         &                      $SB1$ & $43.460_{43.291}^{43.599}$ & $1998.081_{1998.080}^{1998.082}$ & $0.832_{0.832}^{0.833}$ & $0.812_{0.805}^{0.816}$ & $78.987_{78.865}^{79.100}$ & $175.300_{174.989}^{175.519}$ &    $55.033_{54.595}^{55.359}$ & $9.560_{9.555}^{9.564}$ &                         --- &       $5.197_{5.150}^{5.258}$ &                         --- &                       --- \\
          &               $SB1+p(\varpi)$ & $43.423_{43.284}^{43.590}$ & $1998.081_{1998.080}^{1998.082}$ & $0.832_{0.832}^{0.833}$ & $0.810_{0.805}^{0.816}$ & $78.957_{78.871}^{79.097}$ & $175.284_{174.994}^{175.519}$ &    $54.944_{54.568}^{55.340}$ & $9.558_{9.555}^{9.564}$ & $51.674_{51.375}^{51.858}$ &       $5.209_{5.153}^{5.263}$ &   $1.492_{1.460}^{1.541}$ & $0.368_{0.362}^{0.373}$ \\
          &        $SB1+p(m_1|\theta)$ & $43.430_{43.287}^{43.588}$ & $1998.081_{1998.080}^{1998.082}$ & $0.832_{0.832}^{0.833}$ & $0.810_{0.805}^{0.816}$ & $78.981_{78.870}^{79.101}$ & $175.239_{175.004}^{175.514}$ &    $54.951_{54.619}^{55.359}$ & $9.559_{9.555}^{9.564}$ & $55.356_{54.733}^{56.245}$ &       $5.210_{5.150}^{5.256}$ &   $1.182_{1.124}^{1.222}$ & $0.405_{0.399}^{0.414}$ \\
          & $SB1+p(\varpi)+p(m_1|\theta)$ & $43.337_{43.190}^{43.491}$ & $1998.080_{1998.079}^{1998.082}$ & $0.832_{0.832}^{0.832}$ & $0.797_{0.792}^{0.802}$ & $78.879_{78.778}^{79.008}$ & $174.749_{174.475}^{174.962}$ &    $54.046_{53.674}^{54.367}$ & $9.556_{9.552}^{9.561}$ & $52.035_{51.843}^{52.285}$ &       $5.347_{5.299}^{5.399}$ &   $1.380_{1.349}^{1.407}$ & $0.386_{0.381}^{0.391}$ \\
\hline
  36497  & Orb6 & $7.621$ & 2019.554 & 0.692  & 0.102 &  271.8 & 183.6 & 40.4 & --- & --- & --- & --- & --- \\
 & SB9 & $7.415 \pm 0.090$ & $1981.605 \pm 0.078$ & $0.716 \pm 0.098$ & --- & $263.19 \pm 5.24$ & --- & --- & $-2.379 \pm 0.094$ & --- & --- & --- & --- \\
 &                      $SB1$ & $7.645_{7.607}^{7.675}$ & $1981.398_{1981.321}^{1981.495}$ & $0.643_{0.616}^{0.666}$ & $0.093_{0.089}^{0.096}$ &      $265.146_{257.952}^{271.781}$ &    $192.723_{183.966}^{201.801}$ &  $32.491_{26.094}^{37.315}$ & $-2.496_{-2.648}^{-2.282}$ &                         --- & $16.175_{13.290}^{20.638}$ &                         --- &                  --- \\
          &               $SB1+p(\varpi)$ &    $7.641_{7.604}^{7.675}$ & $1981.406_{1981.316}^{1981.497}$ & $0.642_{0.615}^{0.664}$ & $0.092_{0.089}^{0.096}$ &     $264.965_{257.607}^{272.062}$ & $192.611_{183.832}^{202.298}$ &  $31.683_{25.101}^{36.548}$ & $-2.455_{-2.658}^{-2.284}$ &    $21.172_{20.827}^{21.531}$ &   $16.823_{13.633}^{21.304}$ & $0.913_{0.666}^{1.113}$ & $0.553_{0.398}^{0.812}$ \\
          &        $SB1+p(m_1|\theta)$ &    $7.644_{7.607}^{7.677}$ & $1981.408_{1981.319}^{1981.495}$ & $0.639_{0.614}^{0.664}$ & $0.092_{0.088}^{0.096}$ &     $264.960_{257.353}^{271.764}$ &  $192.895_{183.302}^{201.661}$ &  $30.794_{25.422}^{36.855}$ & $-2.500_{-2.659}^{-2.286}$ &    $19.281_{18.047}^{20.734}$ &   $17.167_{13.716}^{21.408}$ &   $1.244_{1.131}^{1.324}$ & $0.495_{0.395}^{0.641}$ \\
          & $SB1+p(\varpi)+p(m_1|\theta)$ &    $7.632_{7.593}^{7.671}$ & $1981.442_{1981.358}^{1981.516}$ & $0.667_{0.649}^{0.684}$ & $0.097_{0.095}^{0.099}$ &     $269.173_{263.164}^{274.364}$ & $188.173_{179.813}^{196.820}$ &  $37.459_{34.716}^{39.810}$ & $-2.426_{-2.628}^{-2.258}$ &    $21.123_{20.742}^{21.437}$ &   $13.386_{12.482}^{14.964}$ &   $1.200_{1.090}^{1.280}$ & $0.394_{0.357}^{0.454}$ \\
\hline
38414 & Orb6 & $7.388 \pm 0.657$ & $2012.067 \pm 0.206$ & $0.380 \pm 0.000$ & $0.0624 \pm 0.0071$ & $170.0 \pm 0.0$ & $90.3 \pm 6.8$ & $86.2 \pm 7.0$ & --- & --- & --- & --- & --- \\
 & SB9 & $6.9925$ & 1905.51 & 0.38 & --- & 170. & --- & --- & 25.4 & --- & --- & --- & --- \\
         &                      $SB1$ & $7.164_{7.114}^{7.224}$ & $1984.310_{1984.013}^{1984.537}$ & $0.553_{0.514}^{0.607}$ & $0.055_{0.052}^{0.058}$ & $204.818_{201.670}^{206.700}$ & $89.931_{87.866}^{91.375}$ &    $75.422_{70.695}^{79.633}$ & $7.337_{3.593}^{11.816}$ &                         --- &       $185.367_{169.270}^{199.997}$ &                         --- &                       --- \\
          &               $SB1+p(\varpi)$ & $7.586_{7.540}^{7.621}$ & $1982.410_{1982.273}^{1982.601}$ & $0.662_{0.625}^{0.724}$ & $0.050_{0.047}^{0.052}$ & $196.810_{192.150}^{201.030}$ & $90.324_{88.135}^{92.329}$ &    $69.065_{61.549}^{74.513}$ & $28.029_{26.712}^{29.008}$ & $8.881_{8.520}^{9.413}$ &       $47.008_{41.607}^{57.014}$ &   $1.754_{1.212}^{1.995}$ & $0.717_{0.615}^{1.000}$ \\
          &        $SB1+p(m_1|\theta)$ & $7.540_{7.441}^{7.606}$ & $1982.637_{1982.338}^{1983.057}$ & $0.672_{0.603}^{0.720}$ & $0.050_{0.047}^{0.052}$ & $198.631_{193.964}^{203.585}$ & $89.770_{87.448}^{91.707}$ &    $68.068_{61.720}^{75.024}$ & $26.403_{23.047}^{28.469}$ & $6.011_{5.487}^{6.328}$ &       $64.513_{46.083}^{89.141}$ &   $6.058_{5.960}^{6.232}$ & $0.633_{0.416}^{1.000}$ \\
          & $SB1+p(\varpi)+p(m_1|\theta)$ & $6.130_{6.097}^{6.161}$ & $1981.852_{1981.756}^{1982.007}$ & $0.497_{0.477}^{0.540}$ & $0.058_{0.055}^{0.059}$ & $157.473_{154.046}^{163.057}$ & $95.118_{92.935}^{97.683}$ &    $79.683_{75.663}^{82.481}$ & $26.411_{25.879}^{27.129}$ & $8.464_{8.066}^{8.722}$ &       $33.337_{30.043}^{37.271}$ &   $6.058_{5.933}^{6.208}$ & $0.393_{0.345}^{0.444}$ \\
\hline
39261 & Orb6 & $6.6504 \pm 0.0089$ & $2001.2281 \pm 0.025$ & $0.706 \pm 0.024$ & $0.055 \pm 0.018$ & $8.3 \pm 1.8$ & $118.3 \pm 20.1$ & $55.4 \pm 2.9$ & --- & --- & --- & --- & --- \\
 & SB9 & $6.6312 \pm 0.0066$ & $1934.781 \pm 0.0392$ & $0.718 \pm 0.012$ & --- & $5.22 \pm 1.64$ & --- & --- & $-2.10 \pm 0.14$ & --- & --- & --- & --- \\
         &                      $SB1$ & $6.626_{6.598}^{6.646}$ & $1981.187_{1981.171}^{1981.205}$ & $0.777_{0.753}^{0.795}$ & $0.052_{0.050}^{0.054}$ & $13.176_{10.426}^{16.252}$ & $113.564_{110.904}^{116.311}$ &    $47.889_{39.145}^{56.542}$ & $-1.781_{-2.026}^{-1.552}$ &                         --- &       $45.872_{38.446}^{53.677}$ &                         --- &                       --- \\
          &               $SB1+p(\varpi)$ & $6.629_{6.601}^{6.648}$ & $1981.189_{1981.171}^{1981.204}$ & $0.777_{0.754}^{0.794}$ & $0.053_{0.051}^{0.054}$ & $13.929_{10.397}^{16.182}$ & $113.970_{111.158}^{116.396}$ &    $51.126_{43.281}^{55.723}$ & $-1.763_{-2.022}^{-1.541}$ & $10.128_{9.799}^{10.643}$ &       $42.696_{39.686}^{49.059}$ &   $1.800_{1.355}^{2.041}$ & $0.762_{0.699}^{0.999}$ \\
          &        $SB1+p(m_1|\theta)$ & $6.627_{6.599}^{6.647}$ & $1981.195_{1981.170}^{1981.204}$ & $0.769_{0.752}^{0.795}$ & $0.053_{0.050}^{0.054}$ & $14.193_{10.157}^{15.980}$ & $113.275_{110.844}^{116.235}$ &    $50.481_{39.031}^{55.789}$ & $-1.721_{-2.025}^{-1.552}$ & $9.534_{8.769}^{9.783}$ &       $42.654_{38.986}^{53.832}$ &   $2.300_{2.144}^{2.459}$ & $0.685_{0.611}^{0.919}$ \\
          & $SB1+p(\varpi)+p(m_1|\theta)$ & $6.635_{6.603}^{6.651}$ & $1981.192_{1981.174}^{1981.210}$ & $0.767_{0.744}^{0.784}$ & $0.054_{0.052}^{0.055}$ & $14.515_{11.403}^{17.334}$ & $114.370_{111.651}^{116.627}$ &    $53.433_{47.021}^{58.896}$ & $-1.761_{-1.983}^{-1.509}$ & $9.790_{9.500}^{10.126}$ &       $40.589_{36.887}^{44.553}$ &   $2.265_{2.097}^{2.398}$ & $0.659_{0.580}^{0.750}$ \\
\hline
\hline
\end{tabular}
}
\tablenotetext{a}{The parameter $f/\varpi$, with units of pc, introduced in \citet{ViMe22}, condenses the pair of parameters $\varpi,q$, since they are not determinable due to the absence of radial velocities for the secondary component as in the case of $SB2$ systems.}
\end{sidewaystable}
\normalsize

\small
\begin{sidewaystable}[ht]
\centering
\caption{Table 2 (contd.). MAP estimates and 95\% HDPIs derived from the marginal PDFs of the orbital parameters for the four orbital solutions, based on our astrometric data, discussed in the first paragraph of Section 3. $m_1$ and $\varpi$ are the priors used for the mass of the primary component and the trigonometric parallax, respectively. In the first two lines (of six entries) for each $SB1$ system, we report the values provided by Orb6 and SB9, preserving the significant figures included in those catalogues as an indication of their precision (continued on next page). \label{tab:SB1/SB1_2}}
\resizebox{\textwidth}{!}{\begin{tabular}{cccccccccccccc}
\hline\hline
     HIP~\# &                     System &                 $P$ &                       $T$ &                     $e$ &              $a$ &                   $\omega$  &                $\Omega$ &                     $i$  &                $V_0$&              $\varpi$ &                $f/\varpi$ &         $m_1$  &                     $q$ \\
 & & [yr] & [yr] &  & [arcsec] &   [\textdegree] &  [\textdegree] & [\textdegree] & [km/s] & [mas] & [pc] &  [M$_\odot]$ & \\
\hline
\hline
40167 & Orb6 & $17.263 \pm 0.032$ & $1997.743 \pm 0.160$ & $0.180 \pm 0.013$ & $0.3592 \pm 0.0058$ & $287.3 \pm 3.6$ & $81.0 \pm 1.9$ & 150.0 & --- & --- & --- & --- & --- \\
 & SB9 & $17.2539 \pm 0.1615$ & $1981.248 \pm 0.3697$ & $0.119 \pm 0.018$ & --- & $307.0 \pm 8.0$ & --- & --- & $-7.93 \pm 0.08$ & --- & --- & --- & --- \\
         &                      $SB1$ & $17.204_{17.185}^{17.220}$ & $1983.188_{1983.074}^{1983.357}$ & $0.103_{0.099}^{0.107}$ & $0.379_{0.377}^{0.381}$ & $343.309_{341.375}^{346.585}$ & $68.351_{67.230}^{69.318}$ &    $143.574_{142.781}^{144.206}$ & $-7.956_{-8.039}^{-7.812}$ &                         --- &       $10.940_{10.398}^{11.377}$ &                         --- &                       --- \\
          &               $SB1+p(\varpi)$ & $17.198_{17.185}^{17.220}$ & $1983.212_{1983.068}^{1983.357}$ & $0.103_{0.100}^{0.107}$ & $0.379_{0.377}^{0.381}$ & $343.552_{341.213}^{346.498}$ & $68.340_{67.207}^{69.299}$ &    $143.455_{142.810}^{144.237}$ & $-7.906_{-8.036}^{-7.814}$ & $40.874_{40.598}^{41.163}$ &       $10.812_{10.422}^{11.395}$ &   $1.502_{1.412}^{1.568}$ & $0.792_{0.739}^{0.869}$ \\
          &        $SB1+p(m_1|\theta)$ & $17.206_{17.185}^{17.219}$ & $1983.247_{1983.072}^{1983.358}$ & $0.104_{0.100}^{0.107}$ & $0.380_{0.377}^{0.381}$ & $344.352_{341.281}^{346.483}$ & $67.951_{67.171}^{69.303}$ &    $143.282_{142.816}^{144.251}$ & $-7.930_{-8.030}^{-7.812}$ & $42.992_{42.000}^{44.095}$ &       $10.845_{10.400}^{11.366}$ &   $1.241_{1.127}^{1.323}$ & $0.873_{0.820}^{0.948}$ \\
          & $SB1+p(\varpi)+p(m_1|\theta)$ & $17.200_{17.185}^{17.219}$ & $1983.159_{1983.016}^{1983.308}$ & $0.101_{0.098}^{0.104}$ & $0.377_{0.375}^{0.379}$ & $342.733_{340.264}^{345.510}$ & $68.620_{67.432}^{69.538}$ &    $144.097_{143.392}^{144.749}$ & $-7.903_{-8.040}^{-7.821}$ & $41.054_{40.802}^{41.332}$ &       $11.317_{11.012}^{11.859}$ &   $1.404_{1.334}^{1.455}$ & $0.868_{0.827}^{0.949}$ \\
\hline
43109 & Orb6 & $15.0507 \pm 0.0064$ & $1991.247 \pm 0.005$ &  $0.6558 \pm 0.0018$ & $0.2547 \pm 0.0009$ & $266.10 \pm 0.27$ & $107.99 \pm 0.35$ & $50.01 \pm 0.27$ & --- & --- & --- & --- & --- \\
 & SB9 & 15.04 & 1915.93 & 0.61 & --- & 86.8 & --- & --- & 36.4 & --- & --- & --- & --- \\
         &                      $SB1$ & $15.073_{15.063}^{15.084}$ & $1900.819_{1900.742}^{1900.878}$ & $0.656_{0.651}^{0.661}$ & $0.253_{0.250}^{0.255}$ & $85.086_{84.537}^{85.742}$ & $288.913_{287.911}^{289.668}$ &    $49.960_{49.281}^{50.614}$ & $35.613_{35.454}^{35.745}$ &                         --- &       $16.337_{15.848}^{16.904}$ &                         --- &                       --- \\
          &               $SB1+p(\varpi)$ & $15.072_{15.064}^{15.085}$ & $1900.823_{1900.740}^{1900.879}$ & $0.655_{0.651}^{0.661}$ & $0.252_{0.250}^{0.255}$ & $85.234_{84.593}^{85.742}$ & $288.685_{287.984}^{289.637}$ &    $49.844_{49.242}^{50.538}$ & $35.567_{35.450}^{35.739}$ & $26.418_{26.245}^{26.637}$ &       $16.480_{15.849}^{16.901}$ &   $2.166_{2.053}^{2.300}$ & $0.771_{0.721}^{0.810}$ \\
          &        $SB1+p(m_1|\theta)$ & $15.076_{15.063}^{15.084}$ & $1900.801_{1900.749}^{1900.889}$ & $0.657_{0.651}^{0.662}$ & $0.253_{0.250}^{0.255}$ & $85.158_{84.492}^{85.693}$ & $288.851_{288.036}^{289.763}$ & $50.005_{49.400}^{50.710}$ & $35.633_{35.492}^{35.778}$ & $30.191_{28.975}^{31.468}$ &  $16.289_{15.738}^{16.769}$ &  $1.315_{1.155}^{1.500}$ & $0.968_{0.913}^{1.000}$ \\
          & $SB1+p(\varpi)+p(m_1|\theta)$ & $15.074_{15.066}^{15.087}$ & $1900.812_{1900.732}^{1900.867}$ & $0.654_{0.648}^{0.658}$ & $0.252_{0.249}^{0.253}$ & $85.265_{84.577}^{85.760}$ & $288.568_{287.856}^{289.542}$ & $49.655_{48.838}^{50.112}$ & $35.526_{35.390}^{35.674}$ & $26.535_{26.318}^{26.707}$ &  $16.667_{16.236}^{17.253}$ & $2.094_{1.965}^{2.188}$ & $0.793_{0.755}^{0.845}$ \\
\hline
  54061  & Orb6 & $44.448 \pm 0.11$ & $2002.170 \pm 0.094$ & $0.4392 \pm 0.004$ & $0.590 \pm 0.026$ & $232.8 \pm 7.9$ & $9.3 \pm 8.2$ & $159.9 \pm 3.5$ & --- & --- & --- & --- & --- \\
 & SB9 & 43.9992 & 1909.90 & 0.35 & --- & 174. & --- & --- & -8.7 & --- & --- & --- & --- \\
 &                      $SB1$ & $44.088_{43.914}^{44.452}$ & $1914.123_{1913.418}^{1914.496}$ & $0.433_{0.430}^{0.437}$ & $0.594_{0.592}^{0.598}$ & $53.716_{43.766}^{61.244}$ & $187.957_{178.369}^{195.301}$ & $165.364_{163.889}^{168.578}$ &   $-9.710_{-9.964}^{-9.449}$ &                          --- &    $12.895_{8.960}^{18.327}$ &                         --- &                     --- \\
          &               $SB1+p(\varpi)$ & $44.170_{43.912}^{44.447}$ & $1913.957_{1913.418}^{1914.492}$ & $0.433_{0.430}^{0.437}$ & $0.595_{0.592}^{0.598}$ & $52.500_{44.253}^{61.605}$ & $186.795_{178.573}^{195.392}$ & $165.438_{163.851}^{168.434}$ &    $-9.715_{-9.961}^{-9.455}$ & $26.666_{25.595}^{27.485}$ &    $12.193_{10.194}^{18.433}$ &   $3.837_{2.803}^{4.366}$ & $0.482_{0.352}^{0.937}$ \\
          &        $SB1+p(m_1|\theta)$ & $44.132_{43.898}^{44.438}$ & $1914.044_{1913.429}^{1914.511}$ & $0.433_{0.429}^{0.437}$ & $0.595_{0.592}^{0.598}$ & $55.901_{43.903}^{61.426}$ & $190.032_{178.211}^{195.216}$ & $165.754_{163.889}^{168.185}$ &    $-9.781_{-9.959}^{-9.449}$ & $31.916_{29.187}^{35.259}$ &    $12.929_{9.978}^{16.685}$ &   $1.952_{1.401}^{2.402}$ & $0.703_{0.521}^{0.999}$ \\
          & $SB1+p(\varpi)+p(m_1|\theta)$ & $44.199_{43.946}^{44.477}$ & $1913.888_{1913.334}^{1914.400}$ & $0.433_{0.430}^{0.437}$ & $0.593_{0.591}^{0.597}$ & $51.525_{41.935}^{60.446}$ & $185.901_{176.651}^{194.695}$ & $167.155_{165.236}^{169.404}$ &    $-9.681_{-9.951}^{-9.429}$ & $27.211_{26.620}^{28.234}$ &    $17.792_{16.002}^{18.731}$ &   $2.736_{2.431}^{2.966}$ & $0.938_{0.801}^{1.000}$ \\
\hline
55642 & Orb6 & $184.4572 \pm 1.3275$ & $1948.3626 \pm 0.2117$ & $0.53689 \pm 0.00459$ & $1.91525 \pm 0.01550$ & $142.829 \pm 0.878$ & $54.376 \pm 0.846$ & $127.642 \pm 0.598$ & --- & --- & --- & --- & --- \\
 & SB9 & 191.996 & 1948.465 & 0.54 & --- & 140.7 & --- & --- & -11.0 & --- & --- & --- & --- \\
         &                      $SB1$ & $189.516_{188.784}^{190.180}$ & $1948.812_{1948.553}^{1949.039}$ & $0.551_{0.547}^{0.553}$ & $1.926_{1.919}^{1.937}$ & $145.134_{144.501}^{145.836}$ & $56.052_{55.712}^{56.616}$ &    $127.234_{126.787}^{127.504}$ & $-11.151_{-11.195}^{-11.108}$ &                         --- &       $10.523_{9.709}^{11.128}$ &                         --- &                       --- \\
          &               $SB1+p(\varpi)$ & $189.419_{188.762}^{190.168}$ & $1948.799_{1948.553}^{1949.043}$ & $0.549_{0.547}^{0.553}$ & $1.930_{1.918}^{1.937}$ & $145.176_{144.537}^{145.864}$ & $56.167_{55.709}^{56.606}$ &    $127.105_{126.791}^{127.501}$ & $-11.142_{-11.198}^{-11.111}$ & $41.726_{40.920}^{42.987}$ &       $10.263_{9.798}^{11.216}$ &   $1.576_{1.355}^{1.689}$ & $0.749_{0.681}^{0.883}$ \\
          &        $SB1+p(m_1|\theta)$ & $189.317_{188.775}^{190.200}$ & $1948.829_{1948.554}^{1949.030}$ & $0.549_{0.547}^{0.553}$ & $1.927_{1.919}^{1.937}$ & $145.311_{144.527}^{145.822}$ & $56.277_{55.714}^{56.608}$ & $127.153_{126.809}^{127.489}$ & $-11.150_{-11.195}^{-11.109}$ & $43.111_{42.185}^{43.931}$ & $10.348_{9.739}^{11.160}$ & $1.379_{1.310}^{1.447}$ & $0.805_{0.742}^{0.903}$ \\
          & $SB1+p(\varpi)+p(m_1|\theta)$ & $189.561_{188.747}^{190.192}$ & $1948.860_{1948.582}^{1949.053}$ & $0.550_{0.547}^{0.553}$ & $1.926_{1.917}^{1.936}$ & $145.345_{144.631}^{145.906}$ & $56.178_{55.729}^{56.612}$ & $127.249_{126.837}^{127.536}$ & $-11.166_{-11.207}^{-11.130}$ & $42.438_{41.949}^{43.259}$ & $10.723_{10.113}^{11.360}$ & $1.417_{1.336}^{1.461}$ & $0.835_{0.768}^{0.921}$ \\
\hline
  67620 & Orb6 & $10.485 \pm 0.06$ & $2009.218 \pm 0.028$ & $0.3462 \pm 0.0080$ & $0.2827 \pm 0.0014$ & $137.5 \pm 1.4$ & $171.3 \pm 0.3$ & $96.4 \pm 0.2$ & --- & --- & --- & --- & --- \\
 & SB9 & $10.4786 \pm 0.019$ & $2009.248 \pm 0.0329$ & $0.3586 \pm 0.0056$ & --- & $140.3 \pm 1.3$ & --- & --- & $5.242 \pm 0.028$ & --- & --- & --- & --- \\
&                      $SB1$ & $10.455_{10.441}^{10.470}$ & $1998.842_{1998.811}^{1998.879}$ & $0.340_{0.336}^{0.346}$ & $0.286_{0.282}^{0.289}$ & $141.958_{140.932}^{143.040}$ & $171.617_{171.021}^{172.308}$ &    $96.031_{95.495}^{96.710}$ &       $5.351_{5.319}^{5.380}$ &                          --- &       $7.175_{7.087}^{7.320}$ &               --- &                   --- \\
          &               $SB1+p(\varpi)$ & $10.456_{10.442}^{10.470}$ & $1998.843_{1998.806}^{1998.876}$ & $0.340_{0.336}^{0.346}$ & $0.286_{0.282}^{0.288}$ & $141.981_{140.945}^{143.062}$ & $171.636_{171.040}^{172.369}$ &    $96.194_{95.511}^{96.724}$ &       $5.356_{5.321}^{5.381}$ & $53.936_{53.196}^{54.546}$ &       $7.175_{7.090}^{7.315}$ &   $0.835_{0.776}^{0.882}$ & $0.631_{0.614}^{0.656}$ \\
          &        $SB1+p(m_1|\theta)$ & $10.453_{10.442}^{10.470}$ & $1998.850_{1998.808}^{1998.877}$ & $0.340_{0.336}^{0.346}$ & $0.285_{0.282}^{0.289}$ & $142.167_{140.904}^{143.030}$ & $171.738_{171.021}^{172.311}$ &    $96.250_{95.506}^{96.691}$ &       $5.348_{5.320}^{5.380}$ & $50.662_{49.249}^{51.705}$ &       $7.182_{7.090}^{7.319}$ &   $1.037_{0.970}^{1.127}$ & $0.572_{0.552}^{0.593}$ \\
          & $SB1+p(\varpi)+p(m_1|\theta)$ & $10.452_{10.437}^{10.466}$ & $1998.842_{1998.806}^{1998.875}$ & $0.339_{0.335}^{0.344}$ & $0.289_{0.285}^{0.291}$ & $141.761_{140.722}^{142.863}$ & $171.539_{171.030}^{172.349}$ &    $96.138_{95.389}^{96.590}$ &       $5.361_{5.322}^{5.382}$ & $53.095_{52.601}^{53.837}$ &       $7.093_{7.005}^{7.217}$ &   $0.917_{0.855}^{0.950}$ & $0.604_{0.592}^{0.626}$ \\
\hline
75695 & Orb6 & $10.5367 \pm 0.0014$ & $1980.473 \pm 0.0038$ & $0.53971 \pm 0.00021$ & $0.204008 \pm 0.000034$ & $180.21 \pm 0.13$ & $148.041 \pm 0.030$ & $111.452 \pm 0.014$ & --- & --- & --- & --- & --- \\
 & SB9 & 10.496 & 1938.197 & 0.41 & --- & 185.4 & --- & --- & -18.0 & --- & --- & --- & --- \\
         &                      $SB1$ & $10.550_{10.546}^{10.553}$ & $1906.630_{1906.600}^{1906.669}$ & $0.524_{0.520}^{0.528}$ & $0.207_{0.206}^{0.208}$ & $180.651_{179.545}^{181.688}$ & $148.315_{147.991}^{148.652}$ &    $110.874_{110.297}^{111.185}$ & $-19.859_{-19.924}^{-19.784}$ &                         --- &       $16.140_{15.891}^{16.385}$ &                         --- &                       --- \\
          &               $SB1+p(\varpi)$ & $10.550_{10.546}^{10.553}$ & $1906.632_{1906.601}^{1906.669}$ & $0.524_{0.520}^{0.528}$ & $0.207_{0.206}^{0.208}$ & $180.739_{179.560}^{181.726}$ & $148.340_{147.983}^{148.645}$ &    $110.668_{110.275}^{111.174}$ & $-19.861_{-19.929}^{-19.787}$ & $28.781_{27.769}^{30.648}$ &       $16.100_{15.879}^{16.380}$ &   $1.798_{1.380}^{2.032}$ & $0.864_{0.809}^{0.984}$ \\
          &        $SB1+p(m_1|\theta)$ & $10.549_{10.546}^{10.553}$ & $1906.636_{1906.600}^{1906.669}$ & $0.524_{0.520}^{0.528}$ & $0.207_{0.206}^{0.208}$ & $180.604_{179.554}^{181.741}$ & $148.313_{147.985}^{148.643}$ & $110.763_{110.281}^{111.186}$ & $-19.874_{-19.927}^{-19.781}$ & $27.937_{27.520}^{28.455}$ & $16.147_{15.878}^{16.373}$ & $2.009_{1.885}^{2.122}$ & $0.822_{0.796}^{0.852}$ \\
          & $SB1+p(\varpi)+p(m_1|\theta)$ & $10.549_{10.546}^{10.553}$ & $1906.632_{1906.601}^{1906.669}$ & $0.525_{0.520}^{0.528}$ & $0.207_{0.206}^{0.208}$ & $180.569_{179.606}^{181.784}$ & $148.321_{147.980}^{148.646}$ & $110.813_{110.257}^{111.141}$ & $-19.856_{-19.930}^{-19.787}$ & $28.066_{27.663}^{28.550}$ & $16.103_{15.860}^{16.360}$ & $1.982_{1.858}^{2.083}$ & $0.825_{0.800}^{0.857}$ \\
\hline
 76031 & Orb6 & $1.713 \pm 0.002$ & $2016.470 \pm 0.010$ & $0.435 \pm 0.017$ & $0.0427 \pm 0.0010$ & $45.8 \pm 2.3$ & $341.9 \pm 2.1$ & 150.0 & --- & --- & --- & --- & --- \\
 & SB9 & $1.6956 \pm 0.0038$ & $2006.180 \pm 0.0093$ & $0.406 \pm 0.014$ & --- & $43.2 \pm 2.6$ & --- & --- & $4.55 \pm 0.09$ & --- & --- & --- & --- \\
 &                      $SB1$ &    $1.697_{1.693}^{1.699}$ & $2002.798_{2002.783}^{2002.826}$ & $0.461_{0.432}^{0.484}$ & $0.039_{0.037}^{0.040}$ &      $44.804_{40.852}^{51.325}$ & $355.676_{352.768}^{360.000}$ &    $170.163_{161.733}^{176.872}$ & $4.614_{4.287}^{4.908}$ &                      --- &       $65.700_{29.739}^{190.218}$ &                       --- &        --- \\
          &               $SB1+p(\varpi)$ &    $1.697_{1.693}^{1.700}$ & $2002.810_{2002.790}^{2002.832}$ & $0.457_{0.440}^{0.493}$ & $0.041_{0.039}^{0.042}$ &      $48.314_{43.471}^{53.893}$ & $356.260_{351.875}^{359.999}$ &    $153.961_{149.958}^{157.493}$ & $4.672_{4.316}^{4.916}$ & $19.262_{17.461}^{20.873}$ &       $24.282_{20.957}^{27.585}$ &   $1.727_{1.239}^{2.316}$ & $0.879_{0.708}^{1.000}$ \\
          &        $SB1+p(m_1|\theta)$ &    $1.696_{1.694}^{1.701}$ & $2002.814_{2002.790}^{2002.834}$ & $0.474_{0.445}^{0.498}$ & $0.040_{0.039}^{0.042}$ &      $49.302_{43.654}^{54.297}$ & $356.416_{351.466}^{360.000}$ &    $151.234_{148.193}^{153.730}$ & $4.650_{4.321}^{4.939}$ & $21.785_{20.739}^{23.088}$ &       $22.151_{19.831}^{23.900}$ &   $1.149_{1.082}^{1.238}$ & $0.933_{0.834}^{1.000}$ \\
          & $SB1+p(\varpi)+p(m_1|\theta)$ &    $1.698_{1.694}^{1.701}$ & $2002.812_{2002.790}^{2002.835}$ & $0.474_{0.445}^{0.498}$ & $0.040_{0.038}^{0.041}$ &      $48.590_{43.531}^{54.227}$ & $355.171_{351.179}^{359.998}$ &    $151.993_{149.757}^{154.062}$ & $4.682_{4.313}^{4.931}$ & $21.176_{20.337}^{22.100}$ &     $23.046_{21.487}^{24.422}$ &   $1.182_{1.101}^{1.249}$ & $0.953_{0.884}^{1.000}$ \\
\hline
78727 & Orb6 & $45.90 \pm 0.60$ & $1997.22 \pm 0.02$ & $0.744 \pm 0.001$ & $0.654 \pm 0.006$ & $163.8 \pm 5.0$ & $25.3 \pm 4.0$ & $34.5 \pm 1.0$ & --- & --- & --- & --- & --- \\
 & SB9 & 44.699 & 1905.39 & 0.75 & --- & 343.6 & --- & --- & -29.4 & --- & --- & --- & --- \\
         &                      $SB1$ & $45.846_{45.817}^{45.909}$ & $1859.564_{1859.365}^{1859.646}$ & $0.746_{0.745}^{0.748}$ & $0.666_{0.665}^{0.668}$ & $155.300_{154.465}^{155.971}$ & $32.841_{32.270}^{33.587}$ &    $34.607_{34.227}^{34.955}$ & $-32.625_{-34.038}^{-31.509}$ &              --- &       $5.750_{1.274}^{9.470}$ &                  --- &                 --- \\
          &               $SB1+p(\varpi)$ & $45.878_{45.819}^{45.911}$ & $1859.467_{1859.371}^{1859.648}$ & $0.747_{0.745}^{0.748}$ & $0.667_{0.665}^{0.668}$ & $155.171_{154.472}^{156.046}$ & $32.937_{32.208}^{33.569}$ &    $34.648_{34.225}^{34.971}$ & $-32.608_{-33.717}^{-31.259}$ & $35.847_{35.251}^{36.492}$ &  $5.957_{2.811}^{10.687}$ &   $2.405_{1.881}^{2.774}$ & $0.272_{0.087}^{0.577}$ \\
          &        $SB1+p(m_1|\theta)$ & $45.863_{45.818}^{45.911}$ & $1859.515_{1859.378}^{1859.660}$ & $0.746_{0.745}^{0.748}$ & $0.667_{0.665}^{0.668}$ & $155.219_{154.435}^{155.973}$ & $32.899_{32.251}^{33.600}$ &    $34.631_{34.229}^{34.972}$ & $-32.589_{-33.775}^{-31.285}$ & $43.083_{39.385}^{45.295}$ &   $5.593_{2.347}^{10.350}$ &   $1.337_{1.303}^{1.398}$ & $0.317_{0.109}^{0.680}$ \\
          & $SB1+p(\varpi)+p(m_1|\theta)$ & $45.855_{45.822}^{45.913}$ & $1859.543_{1859.361}^{1859.638}$ & $0.746_{0.745}^{0.748}$ & $0.666_{0.665}^{0.668}$ & $155.376_{154.496}^{156.028}$ & $32.806_{32.201}^{33.528}$ &    $34.515_{34.145}^{34.901}$ & $-30.595_{-31.813}^{-30.139}$ & $36.960_{36.598}^{37.357}$ &      $13.424_{13.095}^{13.655}$ &   $1.404_{1.363}^{1.446}$ & $0.985_{0.950}^{1.000}$ \\
\hline
\hline
\end{tabular}
}
\end{sidewaystable}
\normalsize

\small
\begin{sidewaystable}[h]
\centering
\caption{Table 2 (contd.). MAP estimates and 95\% HDPIs derived from the marginal PDFs of the orbital parameters for the four orbital solutions, based on our astrometric data, discussed in the first paragraph of Section 3. $m_1$ and $\varpi$ are the priors used for the mass of the primary component and the trigonometric parallax, respectively. In the first two lines (of six entries) for each $SB1$ system, we report the values provided by Orb6 and SB9, preserving the significant figures included in those catalogues as an indication of their precision. \label{tab:SB1/SB1_3}}
\resizebox{\textwidth}{!}{\begin{tabular}{cccccccccccccc}
\hline\hline
     HIP~\# &                     System &                 $P$ &                       $T$ &                     $e$ &              $a$ &                   $\omega$  &                $\Omega$ &                     $i$  &                $V_0$&              $\varpi$ &                $f/\varpi$ &         $m_1$  &                     $q$ \\
 & & [yr] & [yr] &  & [arcsec] &   [\textdegree] &  [\textdegree] & [\textdegree] & [km/s] & [mas] & [pc] &  [M$_\odot]$ & \\
\hline
\hline
93017 & Orb6 & $63.2489 \pm 3.0169$ & $1972.7274 \pm 1.0996$ & $0.21805 \pm 0.05240$ & $1.24739 \pm 0.01764$ & $288.164 \pm 9.910$ & $49.397 \pm 0.773$ & $115.743 \pm 1.814$ & --- & --- & --- & --- & --- \\
 & SB9 & $61.391 \pm Fixed$ & $1972.220$ & 0.25 & --- & 102 & --- & --- & $-45.82 \pm 0.69$ & --- & --- & --- & --- \\
         &                      $SB1$ & $61.285_{61.136}^{61.447}$ & $1910.765_{1910.601}^{1910.916}$ & $0.269_{0.266}^{0.273}$ & $1.272_{1.270}^{1.275}$ & $101.048_{100.639}^{101.432}$ & $48.879_{48.659}^{49.040}$ &    $114.107_{113.980}^{114.237}$ & $-45.947_{-46.060}^{-45.857}$ &              --- &  $5.825_{5.503}^{6.071}$ &                  --- &                 --- \\
          &               $SB1+p(\varpi)$ & $61.291_{61.135}^{61.441}$ & $1910.759_{1910.608}^{1910.914}$ & $0.269_{0.266}^{0.273}$ & $1.272_{1.270}^{1.275}$ & $101.036_{100.642}^{101.433}$ & $48.850_{48.672}^{49.045}$ &    $114.136_{113.987}^{114.241}$ & $-45.961_{-46.060}^{-45.858}$ & $67.125_{66.917}^{67.375}$ &  $5.785_{5.494}^{6.082}$ &   $1.108_{1.067}^{1.148}$ & $0.635_{0.584}^{0.690}$ \\
          &        $SB1+p(m_1|\theta)$ & $61.274_{61.140}^{61.441}$ & $1910.765_{1910.615}^{1910.917}$ & $0.269_{0.266}^{0.273}$ & $1.273_{1.270}^{1.275}$ & $100.962_{100.667}^{101.452}$ & $48.864_{48.658}^{49.032}$ &  $114.097_{113.985}^{114.243}$ & $-45.942_{-46.063}^{-45.860}$ & $65.963_{65.072}^{67.013}$ &  $5.812_{5.507}^{6.068}$ &  $1.180_{1.125}^{1.223}$ & $0.622_{0.577}^{0.661}$ \\
          & $SB1+p(\varpi)+p(m_1|\theta)$ & $61.286_{61.117}^{61.417}$ & $1910.743_{1910.629}^{1910.932}$ & $0.270_{0.267}^{0.273}$ & $1.273_{1.270}^{1.275}$ & $100.938_{100.585}^{101.362}$ & $48.839_{48.673}^{49.045}$ &    $114.072_{113.965}^{114.218}$ & $-45.969_{-46.082}^{-45.891}$ & $67.049_{66.859}^{67.301}$ &      $5.609_{5.384}^{5.852}$ &   $1.138_{1.104}^{1.166}$ & $0.603_{0.567}^{0.648}$ \\
\hline
96302 & Orb6 & 4.56 & 1985.56 & 0.82 & 0.030 & 45.5 & 29.3 & 114.6 & --- & --- & --- & --- & --- \\
 & SB9 & $4.303 \pm 0.0011$ & $1988.771 \pm 0.0011$ & $0.7787 \pm 0.0018$ & --- & $139.6 \pm 0.4$ & --- & --- & $-17.26 \pm 0.05$ & --- & --- & --- & --- \\
         &                      $SB1$ & $4.303_{4.301}^{4.306}$ & $1919.920_{1919.873}^{1919.953}$ & $0.789_{0.784}^{0.794}$ & $0.029_{0.026}^{0.031}$ & $139.671_{138.660}^{140.614}$ & $181.511_{168.698}^{188.989}$ &    $106.470_{96.584}^{123.167}$ & $-17.242_{-17.375}^{-17.093}$ &              --- &  $72.573_{64.879}^{88.625}$ &                  --- &                 --- \\
          &               $SB1+p(\varpi)$ & $4.304_{4.301}^{4.306}$ & $1919.910_{1919.870}^{1919.952}$ & $0.788_{0.783}^{0.793}$ & $0.029_{0.026}^{0.031}$ & $139.491_{138.758}^{140.652}$ & $179.270_{168.968}^{188.189}$ &    $107.302_{97.853}^{122.827}$ & $-17.272_{-17.372}^{-17.085}$ & $5.380_{5.176}^{5.581}$ &  $71.053_{66.395}^{89.369}$ &   $5.389_{2.890}^{6.252}$ & $0.619_{0.539}^{0.913}$ \\
          &        $SB1+p(m_1|\theta)$ & $4.304_{4.301}^{4.306}$ & $1919.904_{1919.870}^{1919.952}$ & $0.788_{0.784}^{0.793}$ & $0.028_{0.027}^{0.031}$ & $139.593_{138.722}^{140.623}$ & $181.684_{174.361}^{190.523}$ &  $104.776_{95.343}^{113.757}$ & $-17.266_{-17.374}^{-17.095}$ & $6.691_{6.175}^{7.516}$ &  $72.257_{65.111}^{79.134}$ &  $2.155_{1.775}^{2.423}$ & $0.936_{0.893}^{1.000}$ \\
          & $SB1+p(\varpi)+p(m_1|\theta)$ & $4.303_{4.301}^{4.307}$ & $1919.922_{1919.869}^{1919.952}$ & $0.789_{0.784}^{0.794}$ & $0.025_{0.024}^{0.026}$ & $140.029_{138.951}^{140.815}$ & $173.956_{168.656}^{179.938}$ &    $116.725_{110.479}^{121.808}$ & $-17.195_{-17.368}^{-17.083}$ & $5.464_{5.307}^{5.684}$ &      $89.468_{84.767}^{93.464}$ &   $2.611_{2.270}^{2.908}$ & $0.956_{0.902}^{1.000}$ \\
\hline
 103655  & Orb6 & 28.90 & 2007.26 & 0.656 & 0.709 & 116.0 & 143.5 & 36.3 & --- & --- & --- & --- & --- \\
 & SB9 & $29.51 \pm 0.66$ & $1976.88 \pm 0.046$ & $0.717776 \pm 0.014$ & --- & $308.961 \pm 3.3$ & --- & --- & $-33.9674 \pm 0.095$ & --- & --- & --- & --- \\
 &                      $SB1$ & $29.466_{29.380}^{29.590}$ & $1947.812_{1947.594}^{1947.976}$ & $0.610_{0.605}^{0.617}$ & $0.674_{0.669}^{0.678}$ &    $326.055_{322.815}^{328.055}$ & $110.038_{108.160}^{113.120}$ & $33.590_{31.994}^{34.754}$ &    $-33.698_{-33.828}^{-33.524}$ &                        --- &       $7.202_{6.729}^{7.688}$ &                       --- &                  --- \\
          &               $SB1+p(\varpi)$ & $29.488_{29.379}^{29.580}$ & $1947.777_{1947.616}^{1947.983}$ & $0.612_{0.606}^{0.616}$ & $0.673_{0.670}^{0.678}$ &    $325.884_{323.178}^{328.224}$ & $110.249_{108.185}^{112.916}$ & $33.621_{32.166}^{34.759}$ &    $-33.651_{-33.805}^{-33.523}$ & $66.536_{66.411}^{66.692}$ &       $7.099_{6.825}^{7.514}$ &   $0.629_{0.590}^{0.661}$ & $0.895_{0.833}^{1.000}$ \\
          &        $SB1+p(m_1|\theta)$ & $29.488_{29.368}^{29.566}$ & $1947.764_{1947.609}^{1947.971}$ & $0.611_{0.604}^{0.615}$ & $0.676_{0.671}^{0.680}$ &    $325.655_{323.925}^{328.803}$ & $110.264_{107.252}^{111.780}$ & $34.273_{33.095}^{35.475}$ &    $-33.600_{-33.712}^{-33.452}$ & $73.521_{71.734}^{75.893}$ &       $6.750_{6.475}^{6.939}$ &   $0.451_{0.413}^{0.484}$ & $0.985_{0.942}^{1.000}$ \\
          & $SB1+p(\varpi)+p(m_1|\theta)$ & $29.557_{29.443}^{29.637}$ & $1947.675_{1947.528}^{1947.882}$ & $0.615_{0.609}^{0.619}$ & $0.668_{0.664}^{0.672}$ &    $324.365_{322.440}^{327.348}$ & $111.956_{109.037}^{113.634}$ & $32.325_{31.251}^{33.379}$ &    $-33.687_{-33.817}^{-33.566}$ & $66.642_{66.465}^{66.751}$ &       $7.464_{7.409}^{7.521}$ &   $0.580_{0.569}^{0.593}$ & $0.990_{0.976}^{1.000}$ \\
\hline
 111685  & Orb6 & $16.77 \pm 0.15$ & $1991.78 \pm 0.08$ & $0.256 \pm 0.009$ & $0.330 \pm 0.003$ & $118.7 \pm 2.2$ & $69.0 \pm 1.0$ & $55.9 \pm 0.6$ & --- & --- & --- & --- & --- \\
 & SB9 & $16.912 \pm 0.0712$ & $2008.719 \pm 0.0601$ & $0.249 \pm 0.007$ & --- & $300.1 \pm 1.3$ & --- & --- & $-58.378 \pm 0.113$ & --- & --- & --- & --- \\
 &                      $SB1$ & $16.748_{16.674}^{16.886}$ & $1991.514_{1991.347}^{1991.621}$ & $0.241_{0.227}^{0.251}$ & $0.344_{0.340}^{0.347}$ &    $295.215_{293.082}^{296.834}$ & $66.478_{65.561}^{67.583}$ & $59.045_{57.954}^{59.804}$ &    $-58.203_{-58.410}^{-57.963}$ &                        --- &       $6.977_{6.399}^{7.564}$ &                       --- &                  --- \\
          &               $SB1+p(\varpi)$ & $16.754_{16.674}^{16.878}$ & $1991.493_{1991.345}^{1991.609}$ & $0.240_{0.227}^{0.250}$ & $0.344_{0.340}^{0.347}$ &    $295.062_{293.103}^{296.814}$ & $66.412_{65.548}^{67.554}$ & $58.929_{57.988}^{59.859}$ &    $-58.148_{-58.397}^{-57.956}$ & $51.432_{48.147}^{54.390}$ &       $6.997_{6.425}^{7.584}$ &   $0.682_{0.537}^{0.851}$ & $0.562_{0.474}^{0.656}$ \\
          &        $SB1+p(m_1|\theta)$ & $16.761_{16.672}^{16.883}$ & $1991.508_{1991.340}^{1991.613}$ & $0.239_{0.228}^{0.251}$ & $0.344_{0.340}^{0.347}$ &    $295.131_{293.108}^{296.878}$ & $66.662_{65.591}^{67.624}$ & $58.889_{57.953}^{59.859}$ &    $-58.208_{-58.400}^{-57.952}$ & $54.998_{53.385}^{56.327}$ &       $6.865_{6.440}^{7.567}$ &   $0.542_{0.500}^{0.577}$ & $0.607_{0.554}^{0.692}$ \\
          & $SB1+p(\varpi)+p(m_1|\theta)$ & $16.802_{16.683}^{16.893}$ & $1991.478_{1991.341}^{1991.614}$ & $0.239_{0.227}^{0.250}$ & $0.342_{0.339}^{0.347}$ &    $295.410_{293.294}^{297.093}$ & $66.556_{65.562}^{67.586}$ & $58.596_{57.835}^{59.668}$ &    $-58.182_{-58.399}^{-57.952}$ & $54.019_{52.817}^{55.465}$ &       $7.137_{6.578}^{7.703}$ &   $0.552_{0.514}^{0.589}$ & $0.627_{0.564}^{0.703}$ \\
\hline
 111974  & Orb6 & $20.829 \pm 0.0030$ & $1983.537 \pm 0.0033$ & $0.73499 \pm 0.00014$ & $0.28798 \pm 0.000049$ & $22.31 \pm 0.12$ & $251.540 \pm 0.076$ & $139.861 \pm 0.032$ & --- & --- & --- & --- & --- \\
 & SB9 & 20.930 & 1983.570 & 0.72 & --- & 204.2 & --- & --- & -10.5 & --- & --- & --- & --- \\
 &                      $SB1$ & $20.837_{20.826}^{20.851}$ & $1900.181_{1900.120}^{1900.226}$ & $0.733_{0.730}^{0.735}$ & $0.290_{0.289}^{0.291}$ &    $22.007_{21.149}^{22.963}$ & $251.203_{250.509}^{252.024}$ & $139.355_{138.720}^{139.937}$ &    $-12.191_{-12.346}^{-11.989}$ &                        --- &       $17.916_{17.449}^{18.324}$ &                       --- &                  --- \\
          &               $SB1+p(\varpi)$ & $20.837_{20.826}^{20.849}$ & $1900.177_{1900.127}^{1900.230}$ & $0.732_{0.730}^{0.735}$ & $0.290_{0.289}^{0.291}$ &    $21.893_{21.073}^{22.886}$ & $251.118_{250.383}^{251.873}$ & $139.178_{138.544}^{139.707}$ &    $-12.062_{-12.263}^{-11.923}$ & $27.986_{27.331}^{28.785}$ &       $17.604_{17.236}^{18.028}$ &   $1.300_{1.177}^{1.397}$ & $0.971_{0.948}^{1.000}$ \\
          &        $SB1+p(m_1|\theta)$ & $20.838_{20.824}^{20.849}$ & $1900.175_{1900.132}^{1900.236}$ & $0.732_{0.730}^{0.735}$ & $0.290_{0.289}^{0.291}$ &    $21.671_{20.866}^{22.656}$ & $250.820_{250.197}^{251.640}$ & $138.771_{138.385}^{139.523}$ &    $-12.006_{-12.102}^{-11.801}$ & $28.876_{28.425}^{29.399}$ &       $17.254_{16.948}^{17.535}$ &   $1.175_{1.112}^{1.227}$ & $0.993_{0.983}^{1.000}$ \\
          & $SB1+p(\varpi)+p(m_1|\theta)$ & $20.836_{20.824}^{20.848}$ & $1900.186_{1900.129}^{1900.235}$ & $0.732_{0.730}^{0.735}$ & $0.290_{0.289}^{0.291}$ &    $21.812_{20.829}^{22.626}$ & $250.946_{250.194}^{251.645}$ & $138.945_{138.403}^{139.530}$ &    $-11.960_{-12.088}^{-11.802}$ & $29.005_{28.509}^{29.433}$ &       $17.210_{16.927}^{17.494}$ &   $1.158_{1.106}^{1.213}$ & $0.997_{0.983}^{1.000}$ \\
\hline
 116259  & Orb6 & $15.70 \pm 0.23$ & $2005.49 \pm 0.01$ & $0.536 \pm 0.007$ & $0.220 \pm 0.002$ & $89.5 \pm 0.8$ & $141.5 \pm 0.3$ & $75.1 \pm 0.4$ & --- & --- & --- & --- & --- \\
 & SB9 & $16.654 \pm 0.354$ & $1989.325 \pm 0.058$ & $0.521 \pm 0.012$ & --- & $97.0 \pm 2.7$ & --- & --- & $-3.359 \pm 0.098$ & --- & --- & --- & --- \\
 &                      $SB1$ & $16.352_{16.285}^{16.392}$ & $1989.136_{1989.102}^{1989.180}$ & $0.537_{0.525}^{0.549}$ & $0.223_{0.220}^{0.226}$ &    $87.683_{87.072}^{88.255}$ & $141.842_{141.103}^{142.544}$ & $74.466_{73.904}^{75.045}$ &    $-3.309_{-3.415}^{-3.224}$ &                        --- &       $12.972_{12.452}^{13.303}$ &                       --- &                  --- \\
          &               $SB1+p(\varpi)$ & $16.340_{16.284}^{16.392}$ & $1989.146_{1989.102}^{1989.181}$ & $0.542_{0.525}^{0.550}$ & $0.223_{0.220}^{0.226}$ &    $87.779_{87.037}^{88.251}$ & $141.865_{141.064}^{142.521}$ & $74.587_{73.901}^{75.035}$ &    $-3.330_{-3.418}^{-3.225}$ & $29.154_{28.924}^{29.518}$ &       $12.848_{12.469}^{13.310}$ &   $1.051_{0.969}^{1.103}$ & $0.599_{0.573}^{0.639}$ \\
          &        $SB1+p(m_1|\theta)$ & $16.359_{16.286}^{16.396}$ & $1989.136_{1989.099}^{1989.179}$ & $0.541_{0.525}^{0.549}$ & $0.223_{0.220}^{0.226}$ &    $87.691_{87.057}^{88.212}$ & $142.018_{141.109}^{142.594}$ & $74.431_{73.935}^{75.068}$ &    $-3.298_{-3.414}^{-3.221}$ & $28.528_{27.713}^{29.100}$ &       $12.903_{12.463}^{13.308}$ &   $1.132_{1.068}^{1.226}$ & $0.583_{0.549}^{0.605}$ \\
          & $SB1+p(\varpi)+p(m_1|\theta)$ & $16.349_{16.286}^{16.394}$ & $1989.145_{1989.101}^{1989.179}$ & $0.543_{0.530}^{0.553}$ & $0.225_{0.222}^{0.227}$ &    $87.745_{86.962}^{88.123}$ & $141.788_{141.155}^{142.623}$ & $74.775_{74.129}^{75.183}$ &    $-3.310_{-3.409}^{-3.223}$ & $29.076_{28.828}^{29.382}$ &       $12.672_{12.331}^{13.094}$ &   $1.093_{1.033}^{1.139}$ & $0.583_{0.559}^{0.614}$ \\
\hline
\hline
\end{tabular}
}
\end{sidewaystable}
\normalsize
\FloatBarrier


\FloatBarrier
\begin{figure}[h]
\includegraphics[width=\textwidth]{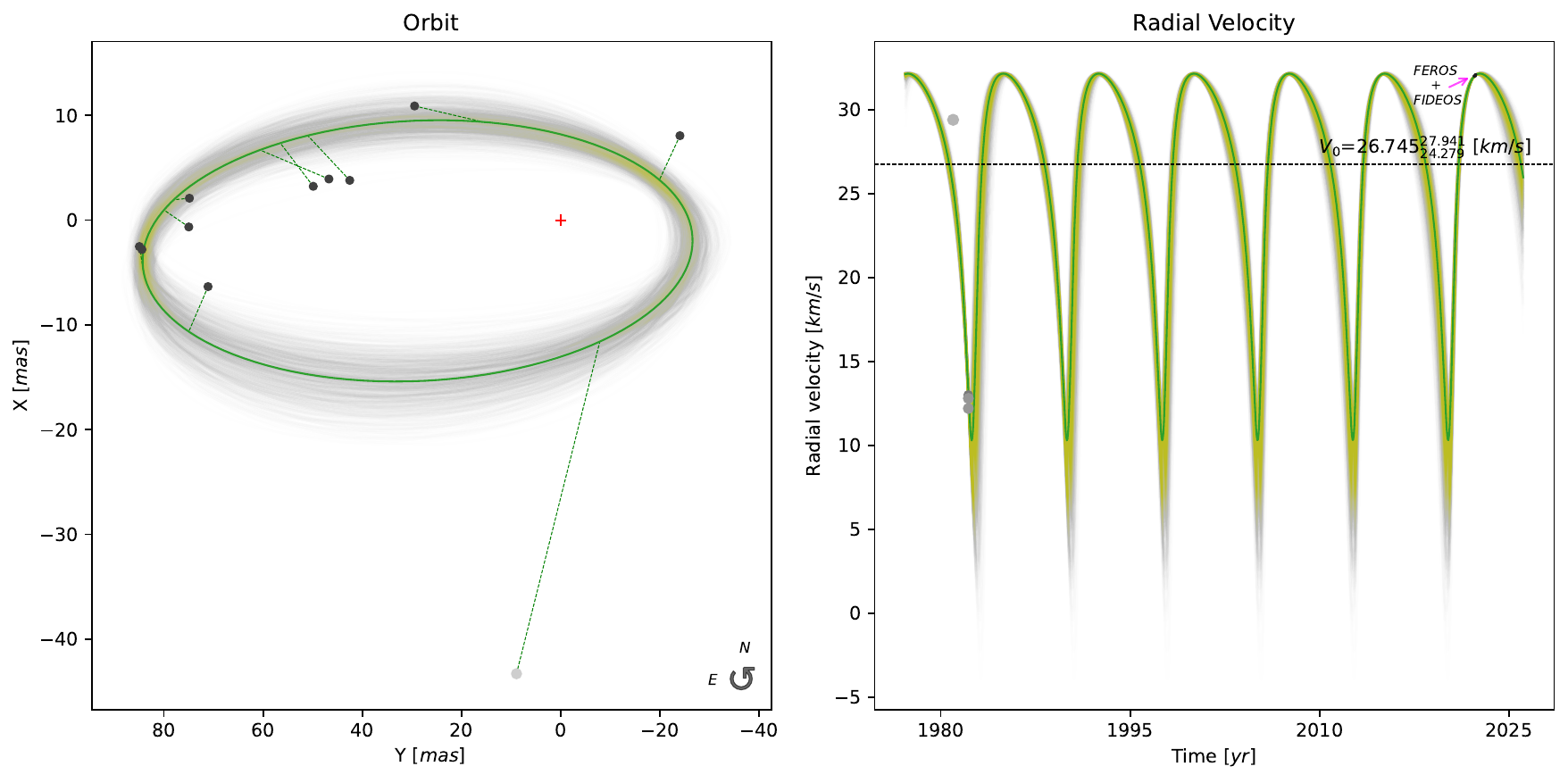}
\caption{Orbit (left panel) and RV curve (right panel) for HIP 38414 based on the MAP values obtained from the $SB1+p(\varpi)+p(m_1|\theta)$ solution given in Table~\ref{tab:SB1/SB1}. The size and color of the dots in both plots depict the weight (uncertainty) of each observation: large clear dots indicate larger errors and the opposite is true for small dark dots. In all astrometric orbits presented, smaller dots are from more recent interferometric measurements, including -but not limited to- our own (although in this particular case, all observations are from SOAR). For this system we have a a phase coverage of about 50\% of the visual orbit. The large deviant point is from SOAR at 2011.9, so gave it a smaller weight in our solution. For the RV curve, we supplemented good quality historical with recent data acquired by us with FEROS and FIDEOS. The dashed horizontal line in the RV curve indicates the estimated systemic velocity, which is included, with its 95\% HDPI range, at the right end of the line. 
\label{fig:38414_3}}
\end{figure}

\begin{figure}[h]
\includegraphics[width=\textwidth]{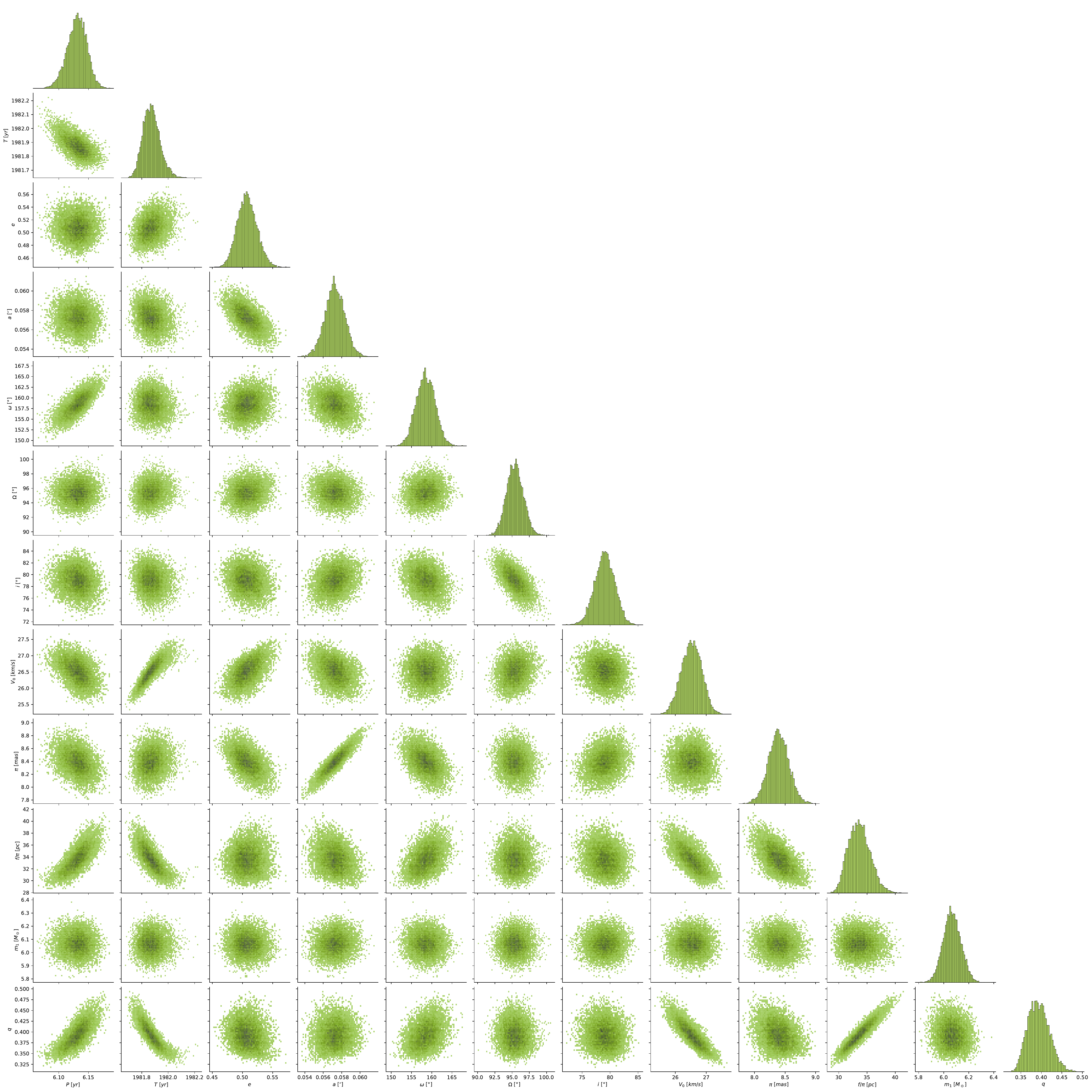}
\caption{Corner plot for HIP 38414. These plots are useful for a qualitative assessment of the quality of the fit, in the sense that better defined orbits, with enough phase coverage, have tight (usually Gaussian-like) PDFs; while less-defined orbits have rather disperse, tangled and/or asymmetrical PDFs with long-tails. Corner plots can also be used to uncover possible correlations between parameters that, if found to be systematic, can eventually be used to reduced the dimensionality of the inference. This is specially useful in problems of high dimensionality. In some cases, we have used these corner plots to check the consistency of our solutions when the SpTy is ambiguous (see the case of HIP~96302, Section~\ref{sec:objnotes}). \label{fig:38414_2}}
\end{figure}

\begin{figure}[h]
\includegraphics[width=\textwidth]{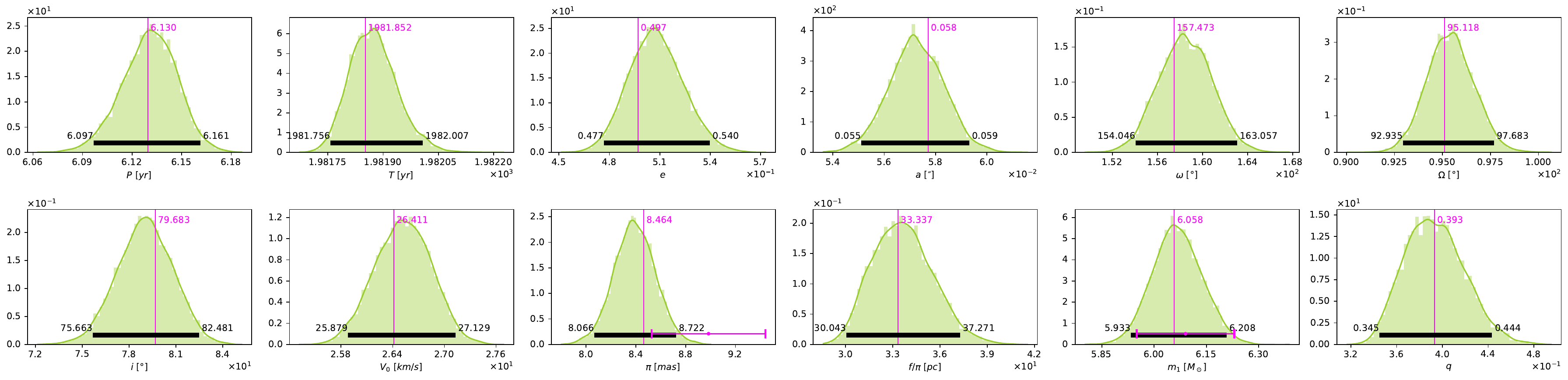}
\caption{Marginal PDFs and MAP estimates (vertical magenta line) for the orbital and physical parameters of the HIP 38414 binary system, for the $SB1+p(\varpi)+p(m_1|\theta)$ solution. The magenta horizontal error bars ($\pm 2\sigma$) indicate the priors adopted for $\varpi$ and  $m_1$, from Table~\ref{tab:SB1/SB1_plx_m1}.
\label{fig:38414_1}}
\end{figure}


\begin{figure}[h]
\includegraphics[width=\textwidth]{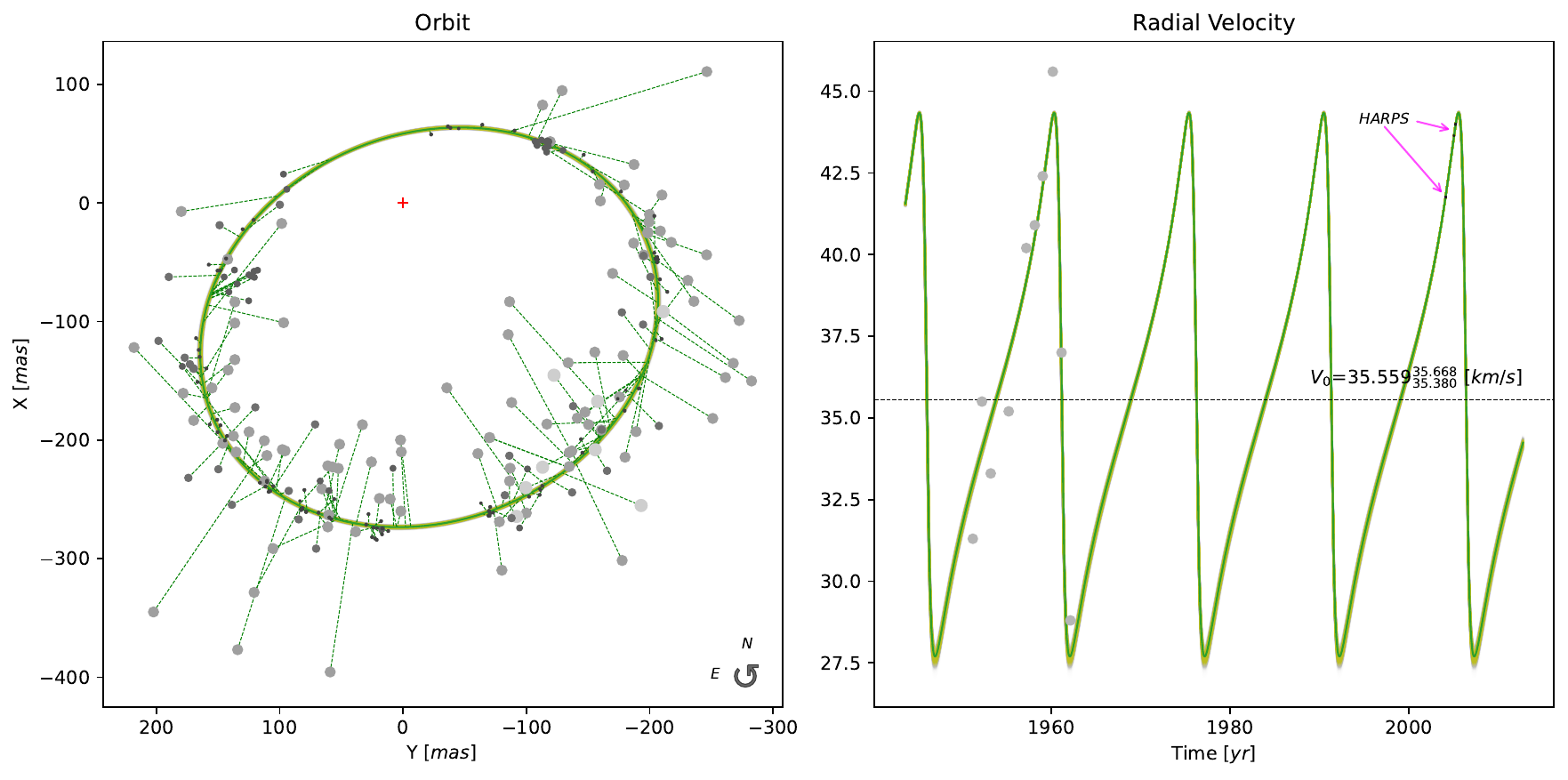}
\caption{Similar to Figure~\ref{fig:38414_3} but for the HIP 43109 system. In this case we have a good orbital coverage of the visual orbit (save for a small arc near periastron). Data points included are of different quality; some are historical RVs of decent quality, and some are highly precise measurements at three consecutive epochs from HARPS at the ESO/La Silla 3.6m telescope.
\label{fig:43109_3}}
\end{figure}


\begin{figure}[h]
\includegraphics[width=\textwidth]{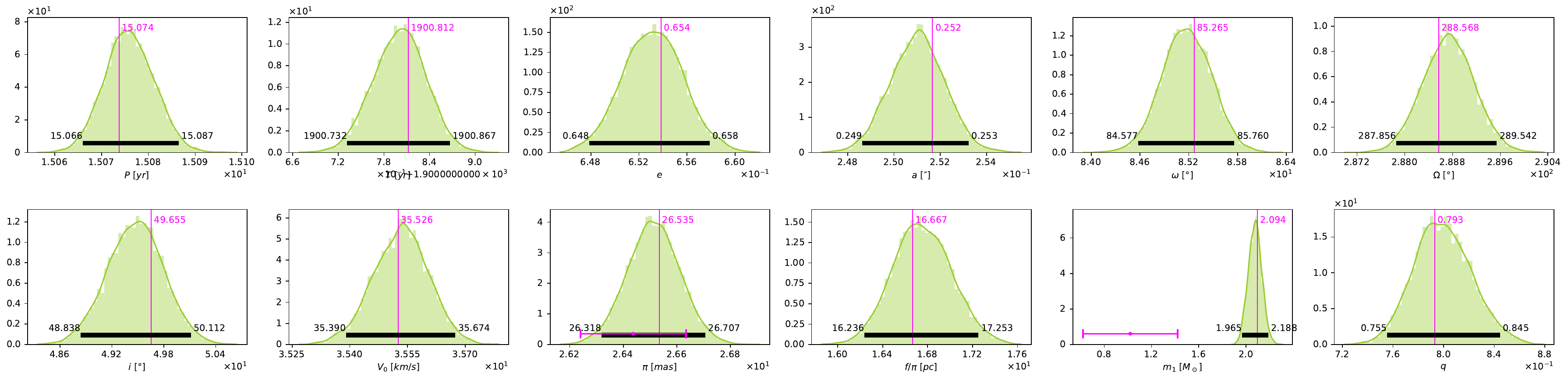}
\caption{Similar to Figure~\ref{fig:38414_1} but for the HIP 43109 system.\label{fig:43109_1}}
\end{figure}
\FloatBarrier

Figures~\ref{fig:38414_3} to 
\ref{fig:43109_1} show the results for two representative $SB1$ systems studied here: HIP 38414 (TOK195) and HIP 43109 (SP1AB). The corresponding figures for all other $SB1s$ presented in this paper can be found at: \url{http://www.das.uchile.cl/~rmendez/B_Research/JAA_RAM_SB1/}\footnote{In this site we also have the data used for our orbital solutions (astrometry and RVs) and their adopted errors. Our own Speckle observations are indicated as SOAR, while our RVs are indicated as FEROS or FIDEOS}. In each case we have produced graphical results of our simultaneous fit to the astrometric and RV data (for the $SB1+p(\varpi)+p(m_1|\theta)$ solution), corner plots, and PDFs for the orbital and physical parameters.

In \citet{ViMe22} a thorough experimental validation of the relative merits of each of the solutions was presented, depending on the prior used (see in particular Sections~3.1.4 and 3.2.4 of that paper). There we conclude that the joint estimation of the orbit and RV curves, subject to the dynamical constraints of the Keplerian motion, allows to share the knowledge provided by both sources of information (trigonometric parallax and SpTy of the primary), reducing the uncertainty of the estimated orbital elements significantly, even if one source of information is highly noisy. Furthermore, we also show that the most robust estimate of the mass ratio is that obtained when both priors ($SB1+p(\varpi)+p(m_1|\theta)$) are used simultaneously. This is true even when there are relatively large differences in the solutions when using different priors, indicative that either the SpTy or the parallax may be somewhat in error or biased.

In Table~\ref{table_masas} we present a global summary of our mass determination for primaries and secondaries based on the data on Table~\ref{tab:SB1/SB1} and the respective tables in \citet{ViMe22} (we adopt the combined solution using both priors, for the reasons explained in the previous paragraph). In Table~\ref{table_masas}, the upper ($m_2^+$) and lower ($m_2^-$) mass for the secondary have been computed from $q^+ \cdot m_1^+$ and $q^- \cdot m_1^-$ respectively, where the $+$ (or $-$) indicates the upper (lower) value of the respective quantities from Table~\ref{tab:SB1/SB1} and its extension (in a way, this is the worst-case scenario for the range of predicted values).

\small
\FloatBarrier
\begin{table}
\centering
\caption{Estimated mass of primary and secondary components of the $SB1$ stellar systems presented in this paper, and in \citet{ViMe22}, obtained when both priors ($SB1+p(\varpi)+p(m_1|\theta)$) are used simultaneously. See text for details. \label{tab:SB1/SB1_masas}}
\begin{tabular}{cccc}
\hline\hline
HIP \# & Discovery   & $m_1$  &
$m_2$  \\
       & Designation & [M$_\odot]$ &   [M$_\odot]$     \\
\hline
171 & BU733AB &$0.927_{0.907}^{0.959}$  & $0.721_{0.684}^{0.762}$ \\
3504 & NOI3Aa,Ab  &$5.765_{3.812}^{7.726}$  & $4.658_{2.695}^{7.726}$ \\
3850 & PES1 &$0.941_{0.864}^{1.012}$  & $0.063_{0.054}^{0.075}$ \\
5336 & WCK1Aa,Ab & $0.921_{0.889}^{0.946}$ & $0.182_{0.165}^{0.212}$  \\
6564 & BU1163  & $1.317_{1.225}^{1.396}$ & $1.275_{1.134}^{1.396}$  \\
7918 & MCY2  & $1.008_{0.981}^{1.035}$ & $0.289_{0.279}^{0.302}$  \\
17491 & BAG8AB  &$0.828_{0.766}^{0.869}$ & $0.565_{0.499}^{0.629}$\\
28691 & MCA24   &$5.334_{3.785}^{6.887}$ & $2.928_{1.805}^{4.580}$ \\
29860 & CAT1Aa,Ab &$1.380_{1.349}^{1.407}$ & $0.533_{0.514}^{0.550}$ \\
36497 & TOK392Da,Db &$1.200_{1.090}^{1.280}$ & $0.473_{0.389}^{0.581}$ \\
38414 & TOK195 &$6.058_{5.933}^{6.208}$ & $2.381_{2.047}^{2.756}$ \\
39261 & MCA33 & $2.265_{2.097}^{2.398}$ & $1.493_{1.216}^{1.799}$ \\
40167 & HUT1Ca,Cb & $1.404_{1.334}^{1.455}$ & $1.219_{1.103}^{1.381}$ \\
43109 & SP1AB & $2.094_{1.965}^{2.188}$ & $1.661_{1.484}^{1.849}$ \\
54061 & BU1077AB & $2.736_{2.431}^{2.966}$ & $2.566_{1.947}^{2.966}$ \\
55642 & STF1536AB & $1.417_{1.336}^{1.461}$ & $1.183_{1.026}^{1.346}$ \\
65982 & HDS1895 & $0.920_{0.878}^{0.955}$ & $0.581_{0.441}^{0.826}$ \\
67620 & WSI77 & $0.917_{0.855}^{0.950}$ & $0.554_{0.506}^{0.595}$ \\
69962 & HDS2016AB & $0.752_{0.700}^{0.814}$ & $0.401_{0.337}^{0.483}$ \\
75695 & JEF1  & $1.982_{1.858}^{2.083}$ & $1.635_{1.486}^{1.785}$ \\
76031 & TOK48  & $1.182_{1.101}^{1.249}$ &$1.126_{0.973}^{1.249}$ \\
78401 & LAB3  & $18.090_{8.753}^{29.606}$ &$10.709_{3.965}^{22.797}$ \\
78727 & STF1998AB & $1.404_{1.363}^{1.446}$ & $1.383_{1.295}^{1.446}$ \\
79101 & NOI2 & $2.350_{2.147}^{2.661}$ & $1.699_{1.168}^{2.661}$ \\
81023 & DSG7Aa,Ab & $1.019_{0.972}^{1.076}$ & $1.013_{0.961}^{1.076}$ \\
93017 & BU648AB & $1.138_{1.104}^{1.166}$ & $0.686_{0.626}^{0.756}$ \\
96302 & WRH32 & $2.611_{2.270}^{2.908}$ & $2.496_{2.048}^{2.908}$ \\
99675 & WRH33Aa,Ab & $9.440_{8.048}^{10.949}$ & $6.476_{5.078}^{9.964}$ \\
103655 & KUI103 & $0.580_{0.569}^{0.593}$ & $0.574_{0.555}^{0.593}$ \\
109951 & HDS3158 & $0.958_{0.934}^{0.990}$ & $0.922_{0.802}^{0.990}$ \\
111685 & HDS3211AB  & $0.552_{0.514}^{0.589}$ & $0.346_{0.290}^{0.414}$ \\
111974 & HO296AB  & $1.158_{1.106}^{1.213}$ & $1.155_{1.087}^{1.213}$ \\
115126 & MCA74Aa,Ab  & $1.195_{1.151}^{1.241}$ & $0.754_{0.708}^{0.804}$ \\
116259 & HDS3356 & $1.093_{1.033}^{1.139}$ & $0.637_{0.577}^{0.699}$ \\
\hline\hline
\end{tabular}
\label{table_masas}
\end{table}
\FloatBarrier
\normalsize

\subsection{A pseudo mass-to-luminosity relationship (MLR) from \textit{SB1}s} \label{pmlr}

As stressed in \citet{ViMe22}, the scheme applied here to $SB1$ systems can only provide informed estimates
of the mass-ratio; definitive values for the individual masses of binary components can still only be obtained in the case of $SB2$ systems. It is interesting however to see how our inferred mass-ratio values compare with a well-defined MLR.

\begin{figure}
\includegraphics[width=18cm,height=14cm]{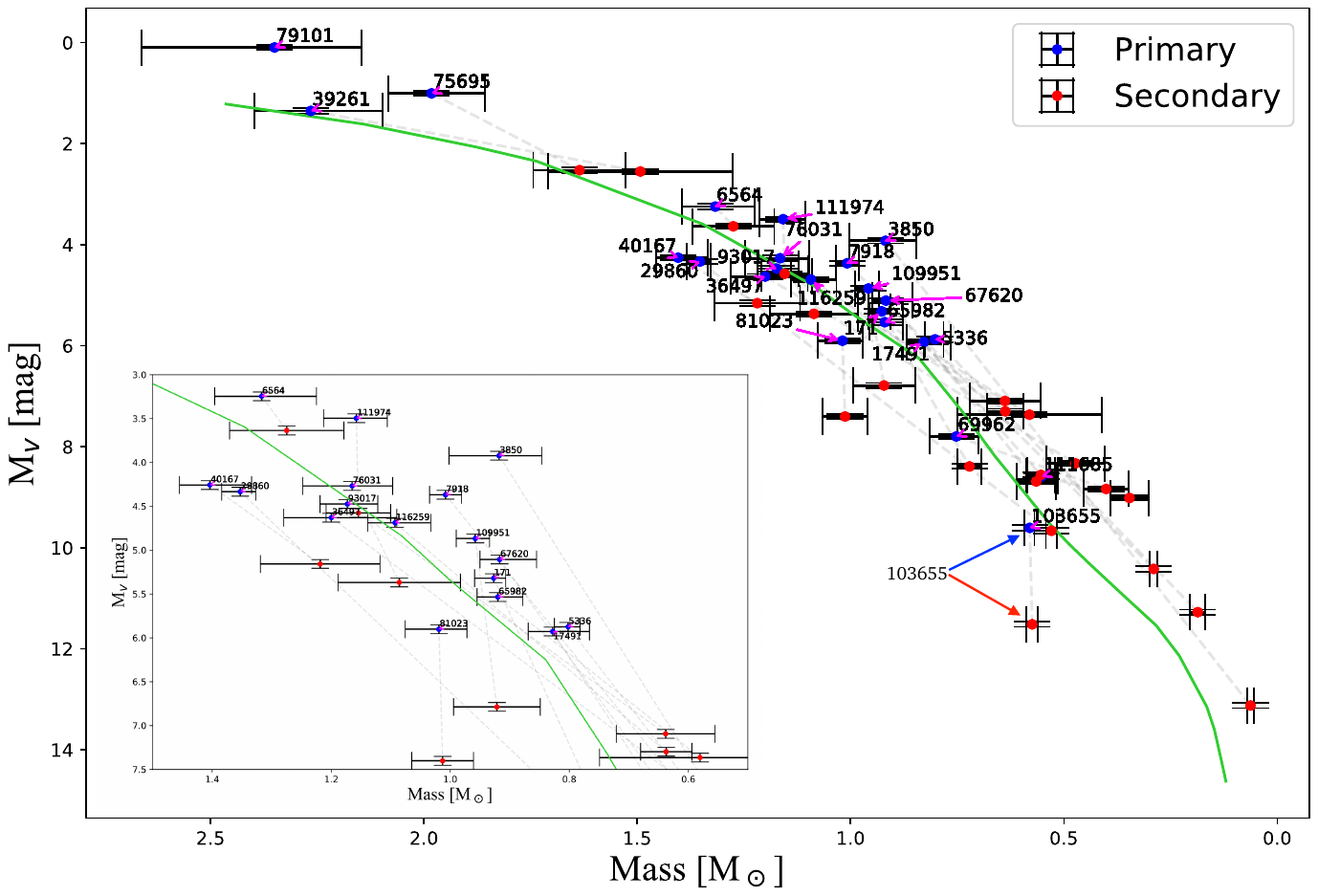}
\caption{Pseudo MLR from the 23 $SB1$ systems of luminosity class V in our sample. The plot inserted shows a zoom in the mass range from 0.5--1.5 M$_\odot$, with $M_V$ from +3.0 -- +7.5. Primary components are depicted with a blue dot, and secondaries with a red dot. The components of each binary are joined by a thin dashed line (the most massive object, HIP 7901 - presented in \citet{ViMe22} - does not have photometry for its secondary, hence only the primary star is shown here). The uncertainty on the mass of the primary and secondary are directly based on the values given in the last row of each entry in Table~\ref{tab:SB1/SB1} (and the corresponding table on \citet{ViMe22}), while we have assumed an uncertainty of $\sim$0.05~mag in $M_V$ as a representative value, considering the errors in the photometry and distance. 
\label{fig:mlr}}
\end{figure}

In Figure~\ref{fig:mlr}, a mass-luminosity plot, we show the position of the fifteen luminosity class V systems among our sample, superimposed to the mean fiducial lines given by \citet{2008PASP..120...38U} (their Figures~5 and~6). In this plot we have also included another eight luminosity class V systems, studied using the same methodology employed here, by \citet{ViMe22}. The results include are those from the solution that provides the lowest uncertainty, which, as mentioned above, is the one that use both priors (SpTy and trigonometric parallax) simultaneously. As can be seen from this figure, there is an overall good agreement between the location of the primaries and our inferred secondaries from the Bayesian statistical method employed here. We note that we do not pretend to build an MLR using these data; this exercise was meant to demonstrate that it is possible to derive tentative mass ratios for $SB1$ systems that could motivate further studies (e.g., to attempt detection/resolution of the secondary, given the $q$ value and the implied luminosity).

\FloatBarrier
\FloatBarrier

\begin{figure}[h]
\includegraphics[width=\textwidth]{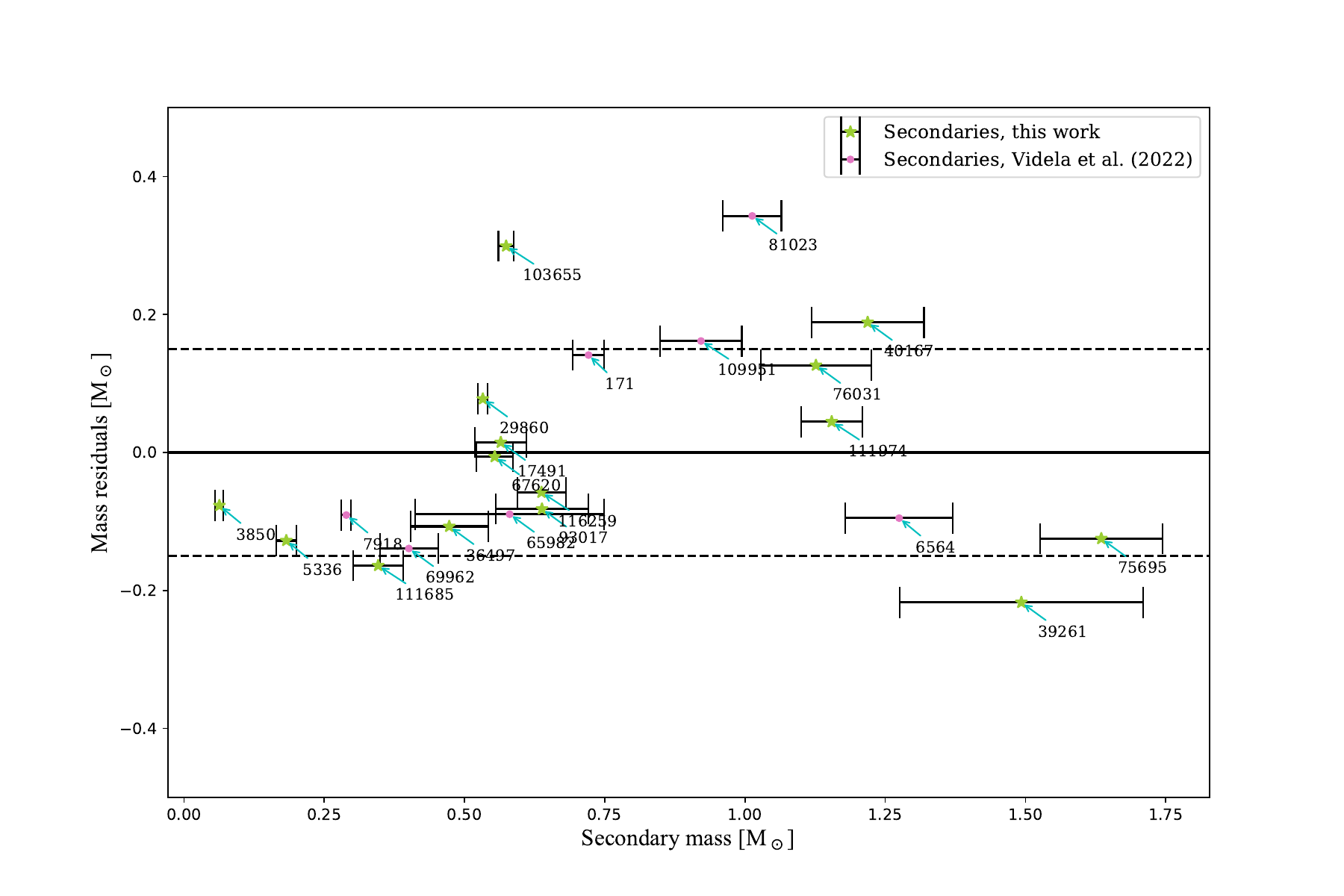}
\caption{Mass residuals for the 22 Class V secondaries of Figure~\ref{fig:mlr} (including those from  \citet{ViMe22}), in comparison with the expected mass given their luminosity from the fidutial relationships from \citet{Abuset2020} (their Table~18). The dashed lines indicate the empirical $\pm 1 \sigma$ value, while the error on mass in the abscissa comes from the upper and lower values presented in Table~\ref{tab:SB1/SB1}. The labels indicate the HIP number.}
\label{fig:mass_comp}
\end{figure}

As it can be seen on Figure~\ref{fig:mlr}, the lateral dispersion (in mass) is reasonable. Indeed, the scatter on mass of our derived secondary masses with respect to a fidutial line is small, amounting to 0.15~$M_\odot$ over the twenty two secondaries plotted in Figure~\ref{fig:mlr}. The mass residuals are shown on Figure~\ref{fig:mass_comp}.

\FloatBarrier
\FloatBarrier

\newpage
\section{Discussion of individual objects}\label{sec:objnotes}

Based on Tokovinin´s Multiple Star Catalogue (MSC\footnote{Updated version available at \url{http://www.ctio.noirlab.edu/~atokovin/stars/}}, \citet{1997A&AS..124...75T, 2018ApJS..235....6T}), twelve of our studied binaries are actually in known multiple systems of different multiplicity, these are HIP 28691, 29860, 36497, 40167, 43109, 54061, 55642, 78727, 93017, 103655, 111685, and 111974, However, since we are looking at the inner (or tighter) components of these systems, their multiplicity does not seem to affect our results, based on an inspection of the residuals of our orbital solutions. The sole exception to this rule is HIP 78727, where we do see extant periodic residuals in position angle and separation, indicative of an unaccounted perturber.
\\
\begin{trivlist}
\item {\bf HIP 3850=PES1:} This system has been extensively studied before by \citet{2019A&A...631A.107P} using spectro-photometry and astrometry, and is a good comparison point for our methodology. The secondary is an L9-type benchmark brown dwarf which leads to the lowest mass-ratio $q$ of our sample (the secondary is the point at the bottom right corner on Figure~\ref{fig:mlr}).

This systems does not have interferometric data, but it has eight high-precision AO observations made with NACO at the ESO/VLT telescope at Cerro Paranal, Chile; covering slightly less than 10~yrs of baseline. The orbital parameters have relatively large errors (specially the Period, with a range from 28--34~yrs), which is a consequence of the small number of observations and the restricted orbit coverage of the available astrometric data. We note that the value of $\omega=-94.2$$^{\circ}$ reported in SB9 is not inconsistent with our value of +264.3$^{\circ}$.

Precise RVs are available from observations made with CORALIE on the Swiss/ESO 1.3m telescope at La Silla, Chile (\citet{2011A&A...525A..95S}), but they cover only 10.3~yrs of the orbit. The SpTy published in WDS is G9V, and G8/K0V in SIMBAD, so we adopted G9V. It is interesting to note that \citet{2019A&A...631A.107P} derive a primary mass of $0.856 \pm 0.014$~$M_\odot$ and a secondary mass of $70.2 \pm 1.6$~$M_{\mbox{Jup}}$ from spectro-photometric data, giving a mass ratio of $q=0.0783 \pm 0.022$, while our best solution (last line on the first row of Table~\ref{tab:SB1/SB1}) gives an inferred value of $q=0.067$, i.e., less than $1\sigma$ of their reported value. This gives a strong support to the adequacy of our methodology. Although its semi-major axis is at the (lower) edge of resolution by Gaia (0.53~arcsec), the large magnitude difference between primary and secondary (9.2~mag) leads to a small RUWE of 0.93, while the Hipparcos and Gaia parallaxes are equal within less than $1\sigma$ (considering the Hipparcos uncertainty). Our systemic RV $V_0 = 9.926 \pm 0.025$~kms$^{-1}$ compares well with the value reported by Gaia\footnote{The RVs in the Gaia catalog result from the average on a variable time window (depending on the number of scans through the source), and covers up to 34 months of observations, \citet{2022arXiv220605902K}.} of $9.71 \pm 0.12$~kms$^{-1}$.\\



\item {\bf HIP 5336=WCK1Aa,Ab:}  This is the system with the second lowest-$q$. There is abundant historical and recent astrometric data covering most of the orbit (including high-resolution imaging secured with the Hubble Space Telescope, spanning almost two decades), as well as good precision RV data. While SB9 indicates that a combined spectroscopic+visual solution has already been obtained by \citet{2015A&A...574A...6A}, the authors do not list all the orbital elements of their orbit (see second line on second row on Table~\ref{tab:SB1/SB1}). On the contrary, our combined MCMC solution seems quite robust, with low formal uncertainties. \citet{2020ApJ...904..112B} have performed a more recent and detailed spectro-photometric and astrometric study of this system, obtaining $0.7440 \pm 0.0122$~$M_\odot$ for the G5V primary and $0.1728 \pm 0.0035$~$M_\odot$ for the dM companion, implying a $q=0.2335 \pm 0.0061$, which compares quite well with our inferred value of $q=0.198 \pm 0.026$ considering our uncertainty. Our inferred mass for the primary ($0.921 \pm 0.032$~$M_\odot$) is smaller than that expected for a G5V used as prior ($1.05 \pm 0.04$~$M_\odot$; see Table~\ref{tab:SB1/SB1_plx_m1}, and the PDF on the web page), and it is more consistent with a G9V. Incidentally, the apparent magnitude for the primary from WDS and the parallax do imply an SpTy of G9-K0V. This object has a very large proper-motion, $(\mu_\alpha \cos \delta, \mu_\delta)= (3468.25\pm 0.35, -1564.94 \pm 0.37$)~mas, and a large negative systemic velocity $V_0=-97.5$~kms$^{-1}$ (see 10th column on Table~\ref{tab:SB1/SB1}), indicative of Halo-like kinematics (the RV given by Gaia is $-97.09 \pm 0.25$~kms$^{-1}$). Indeed, its measured metallicity indicates [Fe/H] $\sim -0.75$, the lowest measured value in our sample. Despite the large magnitude difference between primary and secondary (5.4~mag), it has a large RUWE (7.0). There is a large difference between the HIPPARCOS and Gaia parallaxes (more than 2~mas), probably because it is nearly resolved by Gaia (semi-major axis of 1.0~arcsec), being the nearest object in our sample, at 7.7~pc. There is a difference of $180^\circ$ between the argument of periapsis ($\omega$) and the longitude of the ascending node ($\Omega$) determined by us, and the corresponding values obtained from the astrometry alone (from Orb6). This is a well known ambiguity that can only be resolved by RV data.\\



\item {\bf HIP 17491=BAG8AB:} This object has a pretty good interferometric coverage of the orbit, except near periastron, and high-precision RV observations covering more than one period that were obtained with CORAVEL on the Danish/ESO 1.5m telescope. As a prior for the SpTy, we adopted K0V, from SIMBAD, which is quite close to that given on WDS (G9.5V), and which seemed more adequate given the (Gaia) parallax. There is however a rather large discrepancy between the HIPPARCOS and Gaia DR3 parallaxes ($38.63 \pm0.79$ {\it vs.} $40.33 \pm 0.25$~mas respectively). Our MAP parallax obtained from the combined solution is quite similar to that of Gaia (40.08~mas; see 11th column on last line of third row in Table~\ref{tab:SB1/SB1}), despite the fact that the RUWE for this objects is the second largest of our sample (see Table~\ref{tab:SB1/SB1_plx_m1}). This in principle indicates that the Gaia parallax could be biased, but it may be that the possible bias is being alleviated by the large brightness contrast: the primary has $V=7.9$ while the secondary has $V=10.7$, and hence the photocenter is almost coincident with the primary itself. A combined orbit is also reported by \citet{2002AstL...28..773B}, but it is not included on SB9 (only on Orb) and hence no systemic velocity from this combined fit is available. Our value of $22.31 \pm 0.13$~kms$^{-1}$ is not incompatible with the Gaia at $26.44 \pm 0.61$~kms$^{-1}$, considering the amplitude of the velocity curve (see our web page with figures). Based on the \citet{2002AstL...28..773B} study, \citet{2012A&A...546A..69M} report a $q=0.723 \pm 0.074$ which is within $1\sigma$ of our value ($q=0.682$). 
\\ 


\item {\bf HIP 28691=MCA24:} This is a triple system with an inner binary AaAb, but our analysis refers to the AB system alone (i.e., we used the center-of-mass velocity of the AaAb pair, and the astrometry for AB). It is difficult to observe because the orbit is seen nearly edge-on, and has a small semi-major axis ($a=53$~mas) and a large eccentricity ($e=0.74$). There is no data on the vicinity of periastron, and the astrometric data (including six recent data points, epochs 2015.9 to 2019.1, from our Speckle survey) covers only a small fraction of the orbit. This is compensated, in part, by abundant spectroscopic observations that cover several periods. In WDS its SpTy is listed as an B8III, but in SIMBAD B8V is indicated. Based on the available photometry and trigonometric parallax, we find the primary to be more consistent with a B8III and at a distance of about 262~pc (in agreement with the analysis by \citet{1986AJ.....92.1162F}). This is the most distant system, and the second most massive of our sample. A combined spectroscopic/astrometric solution has already been obtained by \citet{2000AJ....119.2415S} but our new data adds a handful of points that merits a revision of their solution. Independently, \citet{2020AJ....160....7T} published a purely astrometric orbit (listed in the Orb6 line of the corresponding entry in our Table~\ref{tab:SB1/SB1}. As it can be seen from that table, our values for the combined solution in particular for $P$ and $a$ are both slightly smaller than those from \citet{2020AJ....160....7T}, and with slightly larger errors; about halfway from the SB9 values (at least for $P$). Our MAP parallax is found to be 4.1~mas, close to the Gaia parallax of 3.8~mas, and within $1\sigma$ of their uncertainties. In contrast, the Hipparcos parallax for this system is reported to be $4.54 \pm 0.29$~mas, which is probably a biased. The same correction on $\omega$ and $\Omega$ as for HIP~5336 is seen in this system (see Table~\ref{tab:SB1/SB1}). \\

\item {\bf HIP 29860=CAT1Aa,Ab:} This is the first fully combined orbit for the AaAb sub-system (WDS name CAT*1) of this apparently septuple system. It has the largest eccentricity ($e=0.83$) of the objects in our sample. Less than half the orbit is covered, mostly by our own Speckle data secured between 2008 and 2020. Old, low-precision RV data, has been supplement with recent data from our FIDEOS and FEROS monitoring program (with formal uncertainties on the order of 0.01 kms$^{-1}$), which has greatly helped to pin-down the period. Our first attempts to fit the orbit with our astrometric + RV data failed miserably because the RVs published on SB9 were completely off-scale. A careful look at the source of those RVs on \citet{2013AJ....145...41K} shows that some arbitrary zero-point offsets were applied to the old data to conform to their own data. These authors however, were not concerned with the systemic velocity\footnote{Indeed, in the notes on SB9 it says: "No systemic velocity provided in the paper, the value reported and the offset have been supplied by the author directly".}. Specifically, in their Table~3, they indicate offsets of -5.2550 and -14.2000~kms$^{-1}$ applied to data from \citet{2002ApJ...568..352V} and \citet{1986ApJS...62..147B} respectively, in order for these data to conform to theirs. Because our data indicates that RVs from \citet{2013AJ....145...41K}  are completely off, we undid these corrections, applying offsets of +14.2000~kms$^{-1}$ to the data from \citet{2013AJ....145...41K} (effectively putting the RV scale on the zero point of \citet{1986ApJS...62..147B}), and of +8.945~kms$^{-1}$ to the data from \citet{2002ApJ...568..352V}, while not applying any offset to the data from \citet{1986ApJS...62..147B}. These were the historic RVs used for our fits, and available on the data tables in the web page \url{http://www.das.uchile.cl/~rmendez/B_Research/JAA_RAM_SB1/}. The final combined fit to this system is shown in Figure~\ref{fig:29860_3}, which shows the excellent correspondence between the (corrected) RVs from \citet{2002ApJ...568..352V} (epochs 1996 to 2001, near periastron)  and \citet{2013AJ....145...41K} (epochs 2006 to 2009) with our recent data points from FEROS and FIDEOS. In Figure~\ref{fig:29860_1} we show the corresponding PDFs. Our systemic velocity, $9.556 \pm 0.005$~kms$^{-1}$ agrees reasonably well with the Gaia value at $8.70 \pm 0.20$~kms$^{-1}$ (certainly within the RV curve, see right plot on Figure~\ref{fig:29860_3}), giving us further confidence on our zero-point re-normalization procedure.

In the notes of Orb6, it is indicated that the individual masses are $0.96 \pm 0.18$~$M_\odot$ and $0.67 \pm 0.04$$~M_\odot$ from \citet{2006AJ....132.2318C}, derived from ground-based AO observations of the AaAb pair made with the CFH telescope at Mauna-Kea, plus the RVs from \citet{2002ApJ...568..352V} alone. The CFH observations cover however a very short arc (3 years) considering the orbital period (44 years). 
They reported a period of $28.8 \pm 1.1$~yr which is significantly smaller than all values published since then (see Orb6 and SB9 values), including our own fitted value. Their semi-major axis is also smaller, $a= 0.621 \pm 0.019$~arcsec. Our derived primary mass is somewhat larger, at 1.38$M_\odot$, leading to a smaller $q$ ($0.386 \pm 0.005$) than that implied by \citet{2006AJ....132.2318C} ($0.491 \pm 0.064$), albeit within $1.6\sigma$ of their inferred value and errors. We note that our PDFs indicate that the posterior mass for the primary actually tends to be slightly larger than the input a-priory mass for an F9.5V (1.15~$M_\odot$) from \citet{Abuset2020}, while the a-priory and the posterior parallax are, within the errors, commensurable to each other (see Figure~\ref{fig:29860_1}).

\begin{figure}[h]
\includegraphics[width=\textwidth]{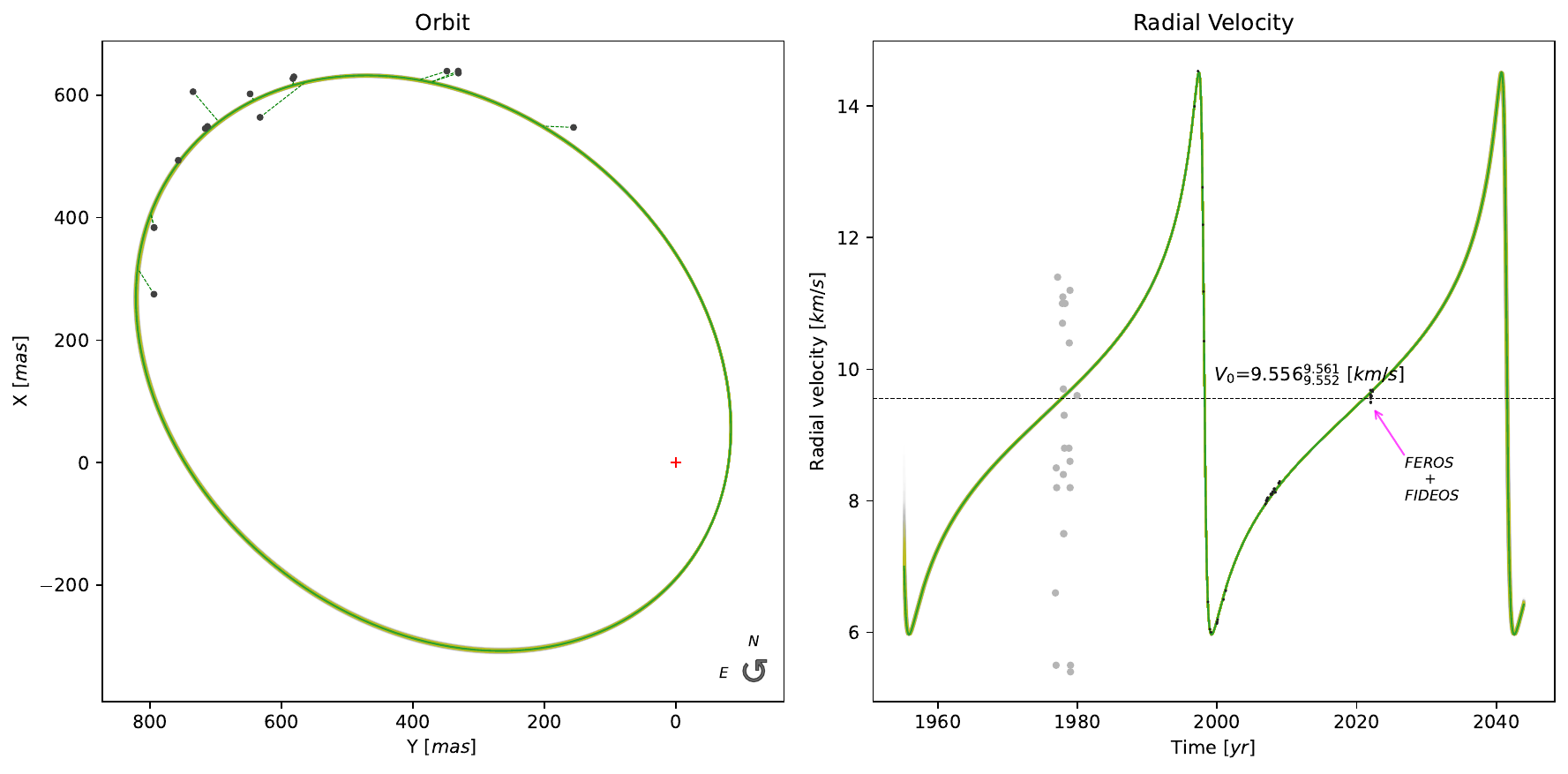}
\caption{Similar to Figure~\ref{fig:38414_3} but for HIP 29860. For this system, the visual orbit is incomplete, with a severe lack of observations near periastron. Old RV data shows a large scatter, while modern data and our own data from FEROS and FIDEOS (obtained in 2021 and 2022 - indicated in the plot), are of much higher quality. This helped to constrain the final orbit which, as indicated in Table~\ref{tab:SB1/SB1}, has a formal uncertainty of 0.3\% in the period,  and 0.6\% in the semi-major axis. \label{fig:29860_3}}
\end{figure}


\begin{figure}[h]
\includegraphics[width=\textwidth]{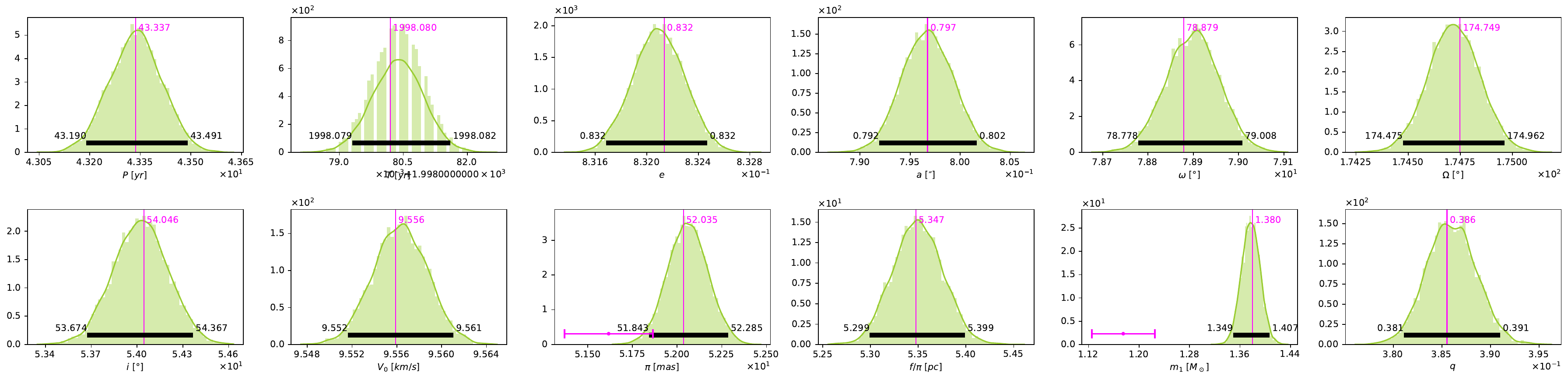}
\caption{Same as Figure~\ref{fig:38414_1} but for the HIP 29860 binary system.\label{fig:29860_1}}
\end{figure}


\item {\bf HIP 36497=TOK392Da,Db:} Ours is the first combined orbit for this $SB1$ binary which is member of a quadruple system. About one-half of the astrometric orbit of the pair is covered by our Speckle data. Previous attempt with the Robo-AO system failed to resolve the binary (\citet{2015ApJ...799....4R}). There is abundant high-quality RV data secured with CORAVEL, covering almost three full periods. There is no ambiguity in the SpTy (F8V). Despite its elevated RUWE (5.80), the Hipparcos, Gaia, and our own MAP parallaxes agree well. Once again, as it was the case for HIP 17491, the primary has $V=8.0~mag$ while the secondary has $V=11.7~mag$, so, the photocenter is almost coincident with the primary itself, which could explain the high RUWE. Our systemic velocity of $-2.43 \pm 0.20$~kms$^{-1}$ is compatible with the Gaia value at $-2.07 \pm 0.43$~kms$^{-1}$.
\\



\item {\bf HIP 38414=TOK195:} Ours is the first combined orbit for this relatively short period $SB1$ system at about 110~pc from the Sun. The K1/2II red giant primary component is the most massive primary object (6.1~$M_\odot$) of our sample. The few historic (from the 1980's) RV points taken from \citet{1983ApJS...53..553P} supplement our own RVs derived from FEROS and FIDEOS observations secured on 2022, which greatly helped our joint solution. The astrometric data covers about one-half of the orbit and is all from our SOAR program. The somewhat elevated residuals are due to the small semi-major axis (62~mas) and the large inclination (almost 80$^{\circ}$; see Figure~\ref{fig:38414_3}\footnote{Indeed in the notes to Orb6 it says: "The binary is difficult to measure, always close to the diffraction limit (on a 4m telescope), and with a magnitude difference $\sim$3."}). While the RUWE value is large (6.0), the difference between the Hipparcos and Gaia parallaxes is small, less than 0.3~mas, within $1\sigma$ of the Gaia uncertainty. Our MAP orbital parallax is $8.46 \pm 0.40$~mas, 0.79~mas smaller than the Gaia parallax, but almost within $1 \sigma$ considering the Gaia error (see Figure~\ref{fig:38414_1}). Finally, it is noteworthy that, given its parallax and photometry, the absolute magnitude ($M_V = -1.0$) is not consistent with a K1.5II star. According to \citet{straivzys1981fundamental}, their Table II, it is about $M_V=-2.5$. We have no explanation for this discrepancy.
\\








\item {\bf HIP 39261=MCA33:} About one-half of the astrometric orbit has been covered for this object and there is abundant RV data of variable precision covering more than two revolutions. According to the notes on Orb6, the mass sum for this system reported in the literature spans a wide range; from  4.1~$M_\odot$ (\citet{1988AN....309...33S}), or $3.61 \pm 0.38$~$M_\odot$ (according to \citet{2004IAUS..224..683B}); to $1.49 \pm 0.66$~$M_\odot$ (from \citet{2002A&A...394..151C}). Our inferred total mass of the system, based on the results of Table~\ref{tab:SB1/SB1}, is $3.73 \pm 0.32$~$M_\odot$ and is within $1\sigma$ of the joint astrometric+spectroscopic solution of \citet{2004IAUS..224..683B}). The PDF for the parallax indicates that our solution has a slightly smaller value than that of Gaia, but within $1.2\sigma$, so our mass estimate does not seem to be affected by the small inferred parallax. The Hipparcos parallax, at $10.13 \pm 0.52$~mas, agrees well with Gaia (despite the elevated RUWE at 5.9) and with our MAP parallax.
\\




\item {\bf HIP 40167=HUT1Ca,Cb:} Ours is the first combined orbit. This $SB1$ system is the CaCb subsystem of a septuple (and, possibly, octuple) system. The coverage of the orbit is good; the last 13 points being from our Speckle survey (2016.9 to 2021.0). The first astrometric measurement is from Hipparcos (1991.25). It also has abundant RV data of reasonable precision covering slightly more than one period. WDS reports a SpTy of M1 (likely referring to the D member; see below), but this does not seem adequate for our object: both, apparent magnitude and distance indicates that the primary (Ca) is a late F (F9; adopted by us). The only paper devoted to this subsystem in particular (see Section~3 in \citet{2000PASP..112..833H}) indicates: ``...thus, we conclude from the color differences that C and D have SpTy G0 and M2, respectively, with an uncertainty only on the order of one spectral subtype.'' Indeed, the photometry for Cb indicates an SpTy of G5-G6V, while the photometry for the D component indicates an M0.
\\





\item {\bf HIP 43109=SP1AB:} This $SB1$ binary is the AB pair ($a=0.25$~arcsec) of a quintuple system. It has a good orbital coverage, including historical data of lower precision and more recent interferometric data (including points in 2001.1, 2014.3, 2018.3, and 2021.2 from our survey), except near periastron. There was no data included in SB9, but we recovered the original RV measurements from  \citet{1939PASP...51..116A} and \citet{1963PDAO...12..159U} which encompass about one full revolution. This system has recently been observed with HARPS (\citet{2020A&A...636A..74T}): \url{https://vizier.u-strasbg.fr/viz-bin/VizieR?-source=J/A+A/636/A74} at 2004.1 (214 measurements), 2005.0 (1 measurement) and 2005.2 (seven measurements)\footnote{Incidentally, in the notes on SB9 it is indicated: ``High-dispersion observations have been continued at Victoria by C.D. Scarfe, and it should be possible soon to give a definitive spectroscopic orbit of this system.''}. The new RVs, with uncertainties below 0.01~kms$^{-1}$, greatly helped constrain the overall fit, which is shown in Figure~\ref{fig:43109_3}. While the published SpTy differ, namely F8V in WDS and G1III in SIMBAD, the apparent magnitude and parallax render it more consistent with the primary being a giant G1III with a mass of 1.02~$M_\odot$ (adopted as prior). However, the MAP mass for the primary from our combined fits leads to a mass that is a factor of two larger, indicating it is a more massive and younger object (see Figure~\ref{fig:43109_1}). Indeed, according to SIMBAD, it is known to be a fast rotator and variable; both characteristics being an indicative of youth. A similar correction on $\omega$ and $\Omega$ with respect to the Orb6 values as that found for HIP~5336 is seen in this  system (see Table~\ref{tab:SB1/SB1_2}).\\

\item {\bf HIP 54061=BU1077AB:} Ours is the first combined orbit. This giant of SpTy G9III has a good orbital coverage, except for a short arc near periastron where the small separation has precluded so far a definitive resolution. The astrometric data includes historical micrometric observations dating back to 1889, as well as interferometric data as recent as 2017. The RVs, which encompass one full orbit, are from the old work by \citet{1937MNRAS..98...92S}, and have a rather large scatter. Initially, it was though to have a very small inclination (fixed at 180$^{\circ}$ in the Hipparcos solution, see \citet{1999A&A...341..121S}), but the inclination is now firmly determined: retrograde, at $167.2 \pm 2.1$$^{\circ}$ (see Table~\ref{tab:SB1/SB1_2}). The PDF for the mass indicates a larger mass (2.7~$M_\odot$) than the input a-priory value (1.93~$M_\odot$ given its SpTy, see Table~\ref{tab:SB1/SB1_plx_m1}). As it can be seen on Table~\ref{tab:SB1/SB1_2} (4th and 5th lines), this could be due to an erroneous parallax. Indeed, this objects does not have a published parallax from Gaia, and the Hipparcos value has a rather large uncertainty. A purely astrometric mass sum of high-quality has been obtained by \citet{2018AJ....155...30B} with the Navy Precision Optical Interferometer, leading to $3.44 \pm 0.11$~$M_\odot$. This value is consistent with our $SB1+p(m_1|\theta)$ solution, which gives $3.32 \pm 0.68$~$M_\odot$, but very far from our $SB1+p(\varpi)$ solution, $5.7 \pm 1.6$~$M_\odot$. This is puzzling considering that they adopted the same Hipparcos parallax. Using both priors simultaneously, we obtained a mass sum of $5.32 \pm 0.57$~$M_\odot$, i.e., $3.3\sigma$ larger than that derived by \citet{2018AJ....155...30B}. Of course, another possibility is that both WDS and SIMBAD are erroneous in the SpTy for the primary. The parallax and photometry indicate an earlier SpTy of B8-B9III, which according \citet{Abuset2020} would imply a mass of around 4~$M_\odot$, which is indeed close to our $SB1+p(\varpi)$ solution which gives 3.8~$M_\odot$ for the primary (see Table~\ref{tab:SB1/SB1_2}). The scarcity and relatively low quality of the available RV data suggests that a better coverage of the RV curve with modern spectrographs, should help solve this puzzle. A similar correction on $\omega$ and $\Omega$ with respect to the Orb6 values as that found for HIP~5336 is seen in this system (see Table~\ref{tab:SB1/SB1_2}).
\\

\item {\bf HIP 55642=STF1536AB:} 
This is the tighter $SB1$ binary -AB- of a triple system. Abundant astrometry of relatively good quality, and covering the whole orbit, is available for this 184+ ~yrs period system (it is the system with the longest period in our sample). Two interferometric observations from SOAR, at epochs 2018.3 and 2021.3, are included. No RV data are given in SB9; only the references (\citet{campbellmoore192855642}, \citet{harper1933}, \citet{petrie1949}, and \citet{1976ApJS...30..273A}) from which we have extracted the RV data. To these, we have added 13 extra recent (2021.5 to 2022.3) high-precision RV measurements obtained with FIDEOS, which fit very well in the RV curve. Our adopted prior parallax is the unweighted average from Gaia DR2 (no data for this object on DR3) for the AB and C components (separation of ~5.5~arcmin). The SpTy reported is F4IV in WDS and F3V in SIMBAD, but the photometry and parallax indicate that the primary is a sub-giant of type F0. We thus adopted the SpTy reported in WDS. Our systemic velocity of $-11.166 \pm 0.041$~kms$^{-1}$ is in reasonable agreement with the Gaia value at $-7.5 \pm 2.7$~kms$^{-1}$, specially considering the large formal uncertainty of the Gaia measurement.
\\

\item {\bf HIP 67620=WSI77:} Pretty good coverage of the astrometric orbit. All but one data point are from our HRCam observations made with SOAR. Almost 1.5 revolutions are covered by the RV curve, including 24 recent data points from HARPS (\citet{2020A&A...636A..74T}), at epochs 2012.2 to 2013.2, which match the orbit very well. The elements given in SB9 are from the spectroscopic-only study by \citet{2016AJ....152...46W}, but there is however a previous combined orbit + RV solution from \citet{2012AJ....144...56T} (given in the Orb6 line of the corresponding row in Table~\ref{tab:SB1/SB1_2}), which compares quite well with our values (see his Table~3). The solution from \citet{2012AJ....144...56T} implies masses of $m_1=0.99$~$M_\odot$ and $m_2=0.63$~$M_\odot$, which are equivalent to our $m_1=0.917 \pm 0.048$~$M_\odot$ and $m_2=0.554 \pm 0.043$~$M_\odot$ given on Table~\ref{tab:SB1/SB1_2}. Our calculated systemic velocity is $5.361 \pm 0.039$~kms$^{-1}$. The Gaia value has a huge error ($17 \pm 23$~kms$^{-1}$) which precludes a proper comparison. The Hipparcos and Gaia DR2 parallax (adopted by us as prior) differ quite substantially, being $53.88 \pm 0.34$~mas and $51.35 \pm 0.4$~mas, respectively. Our inferred MAP value is $53.10 \pm 0.74$~mas, closer to the Hipparcos parallax. There is no parallax on Gaia DR3 for this system (hence no RUWE either).
\\




\item {\bf HIP 75695=JEF1:} Pretty good orbital coverage, including periastron. There are data of various quality, including interferometric measurements, the last of which from our SOAR program (two data points on 2019.1). Historical (1930 to 1943) RV data of good quality, spanning one full orbit, is available from 1930 to 1943, and there also is earlier scattered data from 1902 to 1913  (\citet{1944ApJ....99..134N}), all of which fit the orbit quite well. While there is a parallax from Gaia DR3 ($27.93 \pm 0.97$~mas, albeit with a large RUWE, 7.3) its formal error is 20\% larger than that of Hipparcos ($29.17 \pm 0.76$~mas), so we opted to use the Hipparcos value as a prior. Interestingly enough, our MAP parallax resulted to be $28.07 \pm 0.44$~mas (see Table~\ref{tab:SB1/SB1_2}), within less than $1\sigma$ of the Gaia value (see the PDF in the web page). While SIMBAD indicates a type F2V, and WDS A5V, the photometry and parallax is more consistent with an earlier type, so we adopted WDS's type as a prior. The combined orbit solution by \citet{2010AJ....140.1623M} gives $m_1= 1.71 \pm 0.18$~$M_\odot$ and $m_2= 1.330 \pm 0.074$~$M_\odot$, while we obtain slightly larger masses, $m_1= 1.98 \pm 0.12$~$M_\odot$ and $m_2= 1.63 \pm 0.12$~$M_\odot$ respectively. This is perhaps due to our smaller MAP parallax (they used the Hipparcos value). Of all the objects in our sample, this one has the largest measured metallicity, at  [Fe/H]$~\sim +1$, however its location in the MLR (indicated in Figure~\ref{fig:mlr}) coincides with that of the solar-metallicity mean relationship.
\\

\item {\bf HIP 76031=TOK48:} Ours is the first combined orbit. The orbital coverage of this tight pair ($P=1.7$~yrs, the shortest-period system in our sample) is poor, less than half the orbit. All data available is from our SOAR interferometry and covers epochs from 2009.3 to 2021.3 (15 data points). Due to the small separation, data are lacking near periastron, implying a relatively large uncertainty in the inclination ($i=152.0 \pm 2.2$$^{\circ}$). Fortunately, we have RV of good quality covering several periods, which helps to constrain the fit. We adopted as prior the Hipparcos parallax ($19.67 \pm 0.89$~mas; this is the object in our sample with the largest parallax error, after HIP~111685) due to its smaller formal error (no parallax is given on Gaia DR3, while on DR2 it is $23.56 \pm 1.2$~mas). Our fitted MAP value for the orbital parallax resulted to be $21.18 \pm 0.92$~mas. Our fitted systemic velocity, $4.68 \pm 0.37$~kms$^{-1}$, disagrees slightly with the Gaia value at $6.2 \pm 3.8$~kms$^{-1}$, but note the large formal uncertainty of the Gaia measurement. Also, given the short period of the system, the Gaia value is not incompatible with the RV excursion from -2.5~kms$^{-1}$ to about +16~kms$^{-1}$ seen in the RV curve (see figures in web page).
\\








\item {\bf HIP 78727=STF1998AB:} This $SB1$ binary is the inner system  -AB- of a quintuple system. It has a good coverage of the astrometric orbit since 1825, with data of various quality, including a few data points from our survey at epochs 2008, 2017 and 2019. There are no data on SB9, but we recovered old results from \citet{campbellmoore1928} and \citet{1929ApJ....70..182C}. Unfortunately they cover less than a period and are of relatively low quality, as result of which our fitted value of $V_0$ is rather uncertain. In Gaia DR3 there is a double parallax: based on the coordinate, the first one in the catalog is for the C component and the second one is for the A component (RUWE is nearly 1.3 in both cases not that large). The value in Table~\ref{tab:SB1/SB1_2} refers to the unweighted average of both. As for the SpTy, WDS lists the primary as F5IV, while SIMBAD gives F7V, but the parallax and photometry leads us to believe that the primary is a sub-giant, thus we adopted F5IV as prior.  \citet{2020AJ....159..265T} has reported $m_1 = m_2 = 1.53$~$M_\odot$, which is similar, but slightly larger than our values of $m_1= 1.404 \pm 0.042$~$M_\odot$ and $m_2= 1.383 \pm 0.054$~$M_\odot$. As mentioned in the introduction to this section, this system exhibits periodic residuals after the MCMC orbital fit, most notably in position angle, with a peak-to-peak amplitude of about 14 degrees (see Figure~\ref{fig:78727} top left panel), and a period comparable to that of the system itself ($\sim 50$~yr, top right panel). The trends are less evident in separation (middle panel) or radial velocity (lower panel). It is unlikely that this is a perturbation to the Keplerian orbit induced by the C companion, located almost 8~arcsec away (in comparison with the less than 1~arcsec separation of the AB system), and with an estimated period of more than 1,500~yr, according to Tokovinin´s MSC catalogue. The extant residuals may indicate the presence of an as yet unidentified third body in the AB system itself, an aspect that needs to be further investigated.
\FloatBarrier
\begin{figure}[h]
\includegraphics[width=\textwidth]{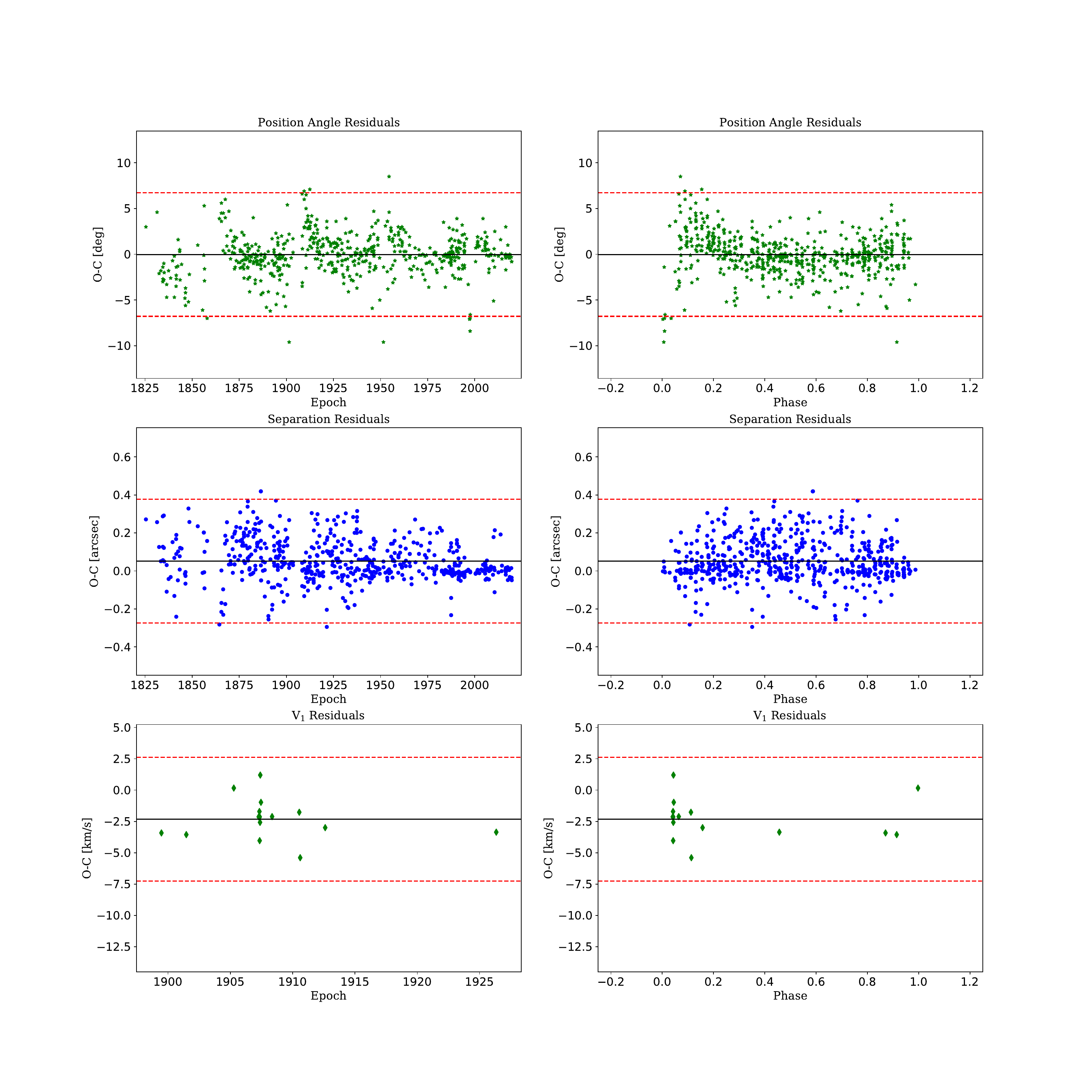}
\caption{Residual O$-$C plots for HIP 78727 based on our MCMC solution with the orbital parameters indicated in Table~\ref{tab:SB1/SB1_2}. The dashed lines indicate the $3\sigma$ boundaries computed from the overall rms on each panel. There is a clear indication of a significant wobble in position angle, with a period of about 50~yr (top panel), possibly due to an accounted companion to the AB system (see text for details). There are hints of some periodicity in the separation residuals as well (middle panel), but less significant. Scarce radial velocity data  precludes us from any conclusion based on the lower panel.
\label{fig:78727}}
\end{figure}
\FloatBarrier

\item {\bf HIP 93017=BU648AB:} Ours is the first combined orbit. At a distance of almost 15~pc, this is the second nearest $SB1$ system in our sample. It is the host of an exoplanet with a 2.8~yrs period. The coverage of the visual orbit is quite complete, including periastron passage, and the data available is in general of good quality. The RVs obtained from the exo-planet campaign (see \citet{1991A&A...248..485D}) cover only a tiny fraction of the orbit (the period of the binary is 61~yrs), but this has been supplemented with newer data in  \citet{2006ApJS..162..207A}, downloaded from \citet{2006yCat..21620207A}, which covers from 2001.5 to 2004.4 (see data tables on web page). There is good correspondence between our systemic velocity, $-45.97 \pm 0.11$~kms$^{-1}$, and that reported by Gaia DR3 at $-43.00 \pm 0.23$~kms$^{-1}$.
\\







 

\item {\bf HIP 96302=WRH32:} Ours is the first combined orbit. This tight binary (separation of 25~mas) is the most compact system in our sample. At nearly 186 pc, it is the second most distant target in our sample. It has scarce astrometric data covering only one-half of the orbit which excludes periastron (where the separation becomes a few mas). It has plenty of RV observations covering several cycles, including old a new higher precision data, hence the period is very well determined. There is a large discrepancy in the published values for the SpTy of the primary: A0V in WDS and G8III in SIMBAD. Its photometry and parallax are incompatible with both of these SpTy, predicting either a B6-7V or a B9III; in both cases much more massive than our inferred value (indeed, see the $SB1+p(\varpi)$ solution in Table~\ref{tab:SB1/SB1_3}). In the end, for our prior, SIMBAD's SpTy was adopted based on the appearance and consistency of our MCMC corner plots for this solution. We note that, with a moderate RUWE of 1.5, the Hipparcos and Gaia values are consistent with each other; $5.37 \pm 0.10$~mas and $5.84 \pm 0.31$~mas respectively, while our MAP orbital parallax is $5.46 \pm 0.22$~mas. The Hipparcos parallax was adopted as prior.

Our inferred mass values are somewhat different than previously published values, with a more massive secondary; \citet{1998A&AS..133..149M} derive $3.344 \pm  1.165$~$M_\odot$ and $1.586 \pm 0.612$~$M_\odot$, whereas we obtain $2.61 \pm  0.32$~$M_\odot$ and $2.50 \pm 0.33$~$M_\odot$ for $m_1$ and $m_2$ respectively. Note the smaller errors of our combined solution. There is good correspondence between our systemic velocity, $-17.20 \pm 0.17$~kms$^{-1}$, and that reported by Gaia DR3 at $-15.8 \pm 1.9$~kms$^{-1}$.
\\





\item {\bf HIP 103655=KUI 103:} This is a triple hierarchical system AaAb,B (AaAb is not resolved). Our solution refers only the AB system, and we treat it as a binary. While AB is considered as $SB2$ in SB9, it has only two measurements of the secondary component. As a result, our MCMC code was unable to converge to a reasonable binary solution, so we decided to treat it as an $SB1$ until more data is secured for the B companion. It has a coverage of about 3/4 of the astrometric orbit with data of reasonable quality. In RV, the phase coverage of the primary is only about one-half the orbit, near periastron. \citet{Pour2000} has published a combined solution, treating it as an $SB2$, but his solution is highly unreliable, the individual derived masses being $3.0 \pm 2.7~M_\odot$ and $1.00 \pm 0.65~M_\odot$. These masses are however incompatible with the SpTy of the primary being M2V, as indicated by both WDS and SIMBAD. These SpTy are furthermore consistent with the apparent magnitude and distance as derived from Gaia DR3 parallax of $66.554 \pm 0.072$~mas (which in turn is also consistent with the Hipparcos parallax at $65.4 \pm 1.8$~mas, despite the large RUWE=5.2). Our derived mass for the primary, $0.580 \pm 0.013~M_\odot$, is somewhat larger than that implied from its SpTy ($0.43~M_\odot$, from \citet{Abuset2020}). The same is true for the secondary, for which $q=0.990 \pm 0.014~M_\odot$, implying  $0.574 \pm 0.015~M_\odot$, while its apparent magnitude and parallax would suggest a SpTy for the secondary of M4-M5, with an implied mass of $0.24-0.31~M_\odot$. We note however that both, WDS and SIMBAD, suggest an earlier type for the secondary, M0.5V (and corresponding mass of about $0.5~M_\odot$), in agreement with our result. This solution however poses a problem, because in the MLR, HIP 103655B is located far below the mean relationship (see Figure~\ref{fig:mlr}), which is because its mass ratio is almost one, but the photometry indicates a $\Delta m\sim 1.9$). We have no explanation for this discrepancy, but, despite these inconsistencies, we can conclude that our solution for this system is more reliable than that presented by \citet{Pour2000}.
\\








\item {\bf HIP 111685=HDS3211AB:} Not a lot of data is available, but it is well spread in both the orbital and RV phase coverage. There is a large discrepancy between the Hipparcos ($51.2 \pm 1.6$~mas) and Gaia DR3 ($46.89 \pm 0.56$ mas) parallaxes. This latter has a very large RUWE (32), which renders the Gaia solution somewhat questionable. Indeed, our solutions are more reliable and consistent adopting as prior the Hipparcos parallax (despite having the largest parallax uncertainty of our sample), leading to a MAP orbital parallax of $54.0 \pm 1.4$~mas about $5\sigma$ larger than the Gaia value. While SB9 reports a combined solution, full orbital parameters are not provided in this catalog (see Table \ref{tab:SB1/SB1_3}).
\\






\item {\bf HIP 111974=HO296AB:} Very good coverage of the visual orbit, with abundant and well spread historical data, as well as newer higher precision data. This includes 20 HRCam data points from our survey, between 2014.76 and 2019.86. No RV data is provided in SB9, so we extracted it from \citet{1985JRASC..79..167B}. We note that while in this paper RV data for the companion is provided (which would place this system in the $SB2$ class), the authors do not use these data and treat it as an $SB1$ (see their Figure~1), probably because of the low precision of these latter data. No Gaia parallax is available, and we have used the Hipparcos value at $29.59 \pm 0.68$ mas as prior, leading to a an inferred MAP orbital parallax of $29.01 \pm 0.50$~mas. We have also treated this system as $SB1$, and it provides an interesting test case of our single-line with priors methodology. \citet{2010AJ....140.1623M} obtained a combined $SB2$ solution for this system using a selected subsample of the \citet{1985JRASC..79..167B} RVs, leading to masses of $1.171 \pm 0.047~M_\odot$ and $1.075 \pm 0.058~M_\odot$, and an (orbital) distance of $34.43 \pm 0.34$~pc. This result compares quite well with our values as can bee seen from Table~\ref{tab:SB1/SB1_3}. In the notes of WDS it says that ``the primary is a giant, from isochrone fit'' (no reference given), while both SIMBAD and WDS indicate a G4V, which is what we have adopted as prior. However, the parallax and photometry are more consistent with an earlier SpTy, F4-5V ($M_V \sim +3.5$), but certainly not a giant.
\\

\item {\bf HIP 116259=HDS3356:} Our is the first combined orbit. This system has a sparse but reasonable coverage of the visual orbit, except near periastron, and abundant good quality RV data covering the full phase space. It has a large RUWE value (8.1), and the Hipparcos ($28.62 \pm 0.95$~mas) and Gaia DR3 ($29.22 \pm 0.15$~mas) parallaxes differ by 0.6~mas. Our MAP orbital parallax is $29.08 \pm 0.31$~mas, i.e., within $1\sigma$ of the Gaia value. There is good correspondence between our systemic velocity, $-3.310 \pm 0.099$~kms$^{-1}$, and that reported by Gaia DR3 at $-1.30 \pm 0.21$~kms$^{-1}$, specially considering that the RV curve has excursions from -10 to +3~kms$^{-1}$ (see plot on web page). It is interesting to note that \citet{2002AJ....124.1144L} obtain a binary mass function of $f(M) = 0.0774 \pm 0.0043~M_\odot$ from RV alone, in perfect agreement with our predicted value from Table~\ref{tab:SB1/SB1_3} of $0.0776~M_\odot$.





\end{trivlist}

\section{Conclusions and final comments} \label{sec:concl}

Applying a Bayesian method developed by our group, we have obtained
mass ratio estimates for 22 $SB1$s with available astrometric and RV
data, using as priors the SpTy of the primary and the trigonometric
parallax of the system. For nine previously unstudied systems, we
present, for the first time, a combined orbital solution and
uncertainty estimates based on a Bayesian approach.  We have made an
exhaustive comparison of our results with previous studies finding a
very good agreement. This includes a comparison of our systemic
velocities with Gaia RVs. We have combined the present results with
those from a previous study by \citet{ViMe22}, for systems of
luminosity class V covering a mass range $0.6 \le M_\odot \le 2.5$ to
construct a pseudo MLR based on 23 systems (45 stars). We find good
correspondence with previously determined MLRs based on $SB2$ systems,
proving the usefulness of our method. Although some inconsistencies
have been found, when the next Gaia data releases are available (with
an improved treatment of binary systems), the parallaxes will become
more reliable and some discrepancies could disappear. An effort is
being made by our team to obtain high signal-to-noise low-resolution
spectra for these (bright) binaries, so that their SpTy and luminosity
class are firmly established. This will open the path to utilize
$SB1$s for more refined studies of the MLR, using larger samples.

\newpage

We would like to thank an anonymous referee as well as the scientific
and statistic editors of the journal for a careful reading of our
manuscript, and for their several very useful comments that
significantly contributed to the readability of our paper.

RAM and EC acknowledge funding from the Vicerrectoria de Investigacion
y Desarrollo (VID) of the Universidad de Chile, project: ENL02/23. RAM
acknowledges the European Southern Observatory in Chile for its
hospitality during a sabbatical leave in which this work was
finished. LV acknowledgs support to FIDEOS operation from
FONDECYT/ANID grant No. 1211162 and Quimal ASTRO20-0025.

We are grateful to Andrei Tokovinin (CTIO/NOIRLab) for his help and support on the use of HRCam at SOAR.
We also acknowledge all the support personnel at CTIO for their commitment to operations during the difficult COVID times.

This research has made use of the Washington Double Star Catalog maintained at the U.S. Naval Observatory and of the SIMBAD database, operated at CDS, Strasbourg. We are very grateful for the continuous support of the Chilean National Time Allocation Committee under programs CN2018A-1, CN2019A-2, CN2019B-13, CN2020A-19, CN2020B-10, CN2021B-17, CN2022A-11, CN2022B-14, and CN2023A-7 for SOAR and CN2022A-1, CN2022B-8, and CN2023A-5 for FEROS.


\vspace{5mm}
\facilities{AURA: SOAR} 


\newpage
\bibliography{sample631}{}
\bibliographystyle{aasjournal}



\end{document}